\documentclass[preprint]{aastex}
\usepackage{graphicx}
\usepackage{placeins}
\usepackage{natbib}
\usepackage{aasref}

\begin{document}

\title{The B-ring's surface mass density from hidden density waves: \\ Less than meets the eye?}
\author{M. M. Hedman$^a$, P.D. Nicholson$^b$}
\affil{$^a$ Department of Physics, University of Idaho, Moscow ID 83844-0903 \\
$^b$ Department of Astronomy, Cornell University, Ithaca NY 14853}

\maketitle

Saturn's B ring is the most opaque ring in our solar system, but many of its fundamental parameters, including its total mass, are not well constrained. Spiral density waves generated by mean-motion resonances with Saturn's moons provide some of the best constraints on the rings' mass density,   but detecting and quantifying such waves in the B ring has been challenging because of this ring's high opacity and abundant fine-scale structure.  Using a wavelet-based analyses of 17 occultations of the star $\gamma$ Crucis observed by the Visual and Infrared Mapping Spectrometer (VIMS) onboard the Cassini spacecraft, we are able to examine five density waves in the B ring. Two of these waves are generated by the Janus 2:1 and Mimas 5:2 Inner Lindblad Resonances at 96,427 km and 101,311 km from Saturn's center, respectively. Both of these waves can be detected in individual occultation profiles, but the multi-profile wavelet analysis reveals unexpected variations in the pattern speed of the Janus 2:1 wave that might arise from the periodic changes in Janus' orbit. The other three wave signatures are associated with the Janus 3:2, Enceladus 3:1 and Pandora 3:2 Inner Lindblad Resonances at 115,959 km, 115,207 km and 108,546 km.  These waves are not visible in individual profiles, but structures with the correct pattern speeds can be detected in appropriately phase-corrected average wavelets. Estimates of the ring's surface mass density derived from these five waves fall between 40 and 140 g/cm$^2$, even though the ring's optical depth in these regions ranges from $\sim 1.5$ to almost 5. This suggests that the total mass of the B ring is most likely between one-third and two-thirds the mass of Saturn's moon Mimas.

\section{Introduction}

The B ring is the brightest, most opaque and probably the most massive of Saturn's rings. It is also among the least well understood, with even basic parameters like its maximum optical depth and its total mass being poorly constrained. Indeed, estimates of the B-ring's mass range from around $1\times10^{19}$ kg \citep{Cooper85} to over $7\times10^{19}$ kg \citep{Robbins10}. The large uncertainties in the B-ring's mass and its typical surface mass density not only hamper efforts to understand the  structure and dynamics of this ring, but also complicate efforts to ascertain the age and history of Saturn's ring system \citep{Charnoz09}. 

\nocite{Shu84}
One reason why the B-ring's mass is so poorly constrained is that very few spiral density or bending waves have been identified in this ring. These spiral patterns provide the most reliable estimates of a ring's mass density, thanks to a very mature theoretical model that describes their formation and propagation \citep{Shu84}. Indeed, analyses of various density waves have yielded numerous mass density estimates of the A ring \citep{Tiscareno07, Tiscareno13}, Cassini Division \citep{Colwell09cd} and C Ring \citep{Baillie11, HN14}.  However, thus far there have only been two waves that have yielded sensible mass density estimates for the B ring. The most prominent wave in the B ring is the one generated by the Janus 2:1 resonance, which lies in the least opaque, innermost part of the B ring. Analyses of Voyager occultation data yielded mass density estimates of around 70 g/cm$^2$ for this region \citep{Holberg82, Esposito83}. On the opposite end of the B ring, \citet{Lissauer85} identified a bending wave due to the 4:2 vertical resonance with Mimas in the Voyager imaging data. Bending waves can be analyzed much like density waves, allowing \citet{Lissauer85} to estimate a local surface mass density of 54$\pm$10 g/cm$^2$. More recently, analyses of normal modes on the B-ring's outer edge have found generally comparable mass densities in this region \citep{SP10, Nicholson14}. 

\begin{figure}[tb]
\resizebox{6in}{!}{\includegraphics{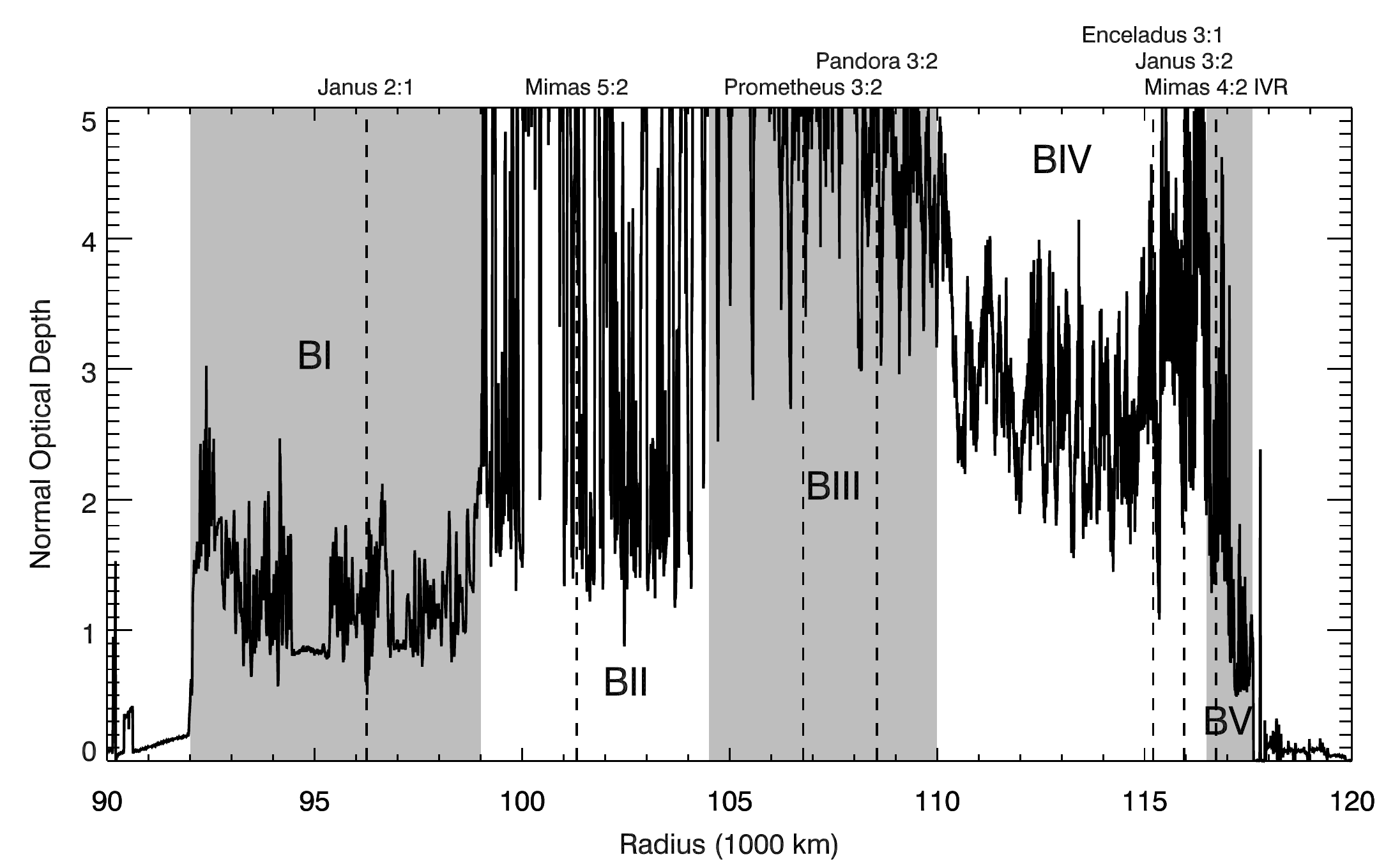}}
\caption{Overview of the B ring.  The profile shows the normal optical depth of Saturn's B ring as a function of ring radius derived from an occultation of the star $\gamma$ Crucis on Cassini Orbit (``Rev'') 089 observed by the VIMS instrument. The background shading indicates the five distinct zones in the B ring \citep{Colwell09}, and the dashed lines mark the locations of mean motion resonances discussed in the text.}
\label{bringov}
\end{figure}

The few existing measurements provide very limited information about the B-ring's surface mass density because they only sample two locations in a very complex ring. As shown in Figure~\ref{bringov}, the B-ring can be divided into five broad regions with very different optical depths \citep{Colwell09}. The Janus 2:1 wave occupies the less opaque BI region, while the Mimas 4:2 wave is found in the BV region, where the ring's structure is highly variable due to its proximity to the highly non-circular outer edge at 117,500 km. To date, no-one has reported mass density estimates for the central  BII, BIII and BIV regions, which include the most opaque parts of Saturn's rings. However, there are several strong satellite resonances in these regions that could generate relatively intense density waves. One is the Mimas 5:2 resonance, which is found in the BII region, where the optical depth appears to switch between values around 2 and near opaque. This wave has been identified in occultation profiles \citep{Colwell13}, but a mass density has not yet been published based on this feature.  In the BIV region, there should be strong resonances with Janus (3:2) and Enceladus (3:1), but neither of these has been identified in the observational data because this part of the B ring contains intense fine-scale stochastic optical depth variations that obscure any organized structure from a density wave. Finally, the Prometheus 3:2 and Pandora 3:2 resonances lie in BIII, the nearly-opaque core of the ring, and so high-resolution occultations, which measure the amount of light transmitted through the rings, have very low signal-to-noise.

Here we investigate all six of these Lindblad resonances in the B ring using a new wavelet-based technique that combines data from multiple occultations in order to identify weak coherent signals due to waves that are not obvious in single measurements. These methods reveal potential wave signatures from the Janus 3:2, Enceladus 3:1 and Pandora 3:2 resonances (but not the Prometheus 3:2 resonance). Together with the Janus 2:1 and Mimas 5:2 density waves and the Mimas 4:2 bending wave, these wave features provide mass density estimates spanning a wide range of optical depths, and thus provide a much clearer picture of the B-ring's total mass. Our measurements indicate that regions with optical depths ranging between 1 and 5 all have mass densities less than 150 g/cm$^2$. This is  consistent with recent investigations of other parts of Saturn's rings, which demonstrate that large variations in the ring's optical depth are often not associated with comparable variations in its surface  mass density \citep{Colwell09cd, Baillie11, Tiscareno13, HN14}.  At the same time, these B-ring mass densities are well below the values that have been considered in some recent studies of the rings' opacity, history and spectral properties \citep{Robbins10, Charnoz11, Hedman13} but may be consistent with estimates based on the rings' charged particle emissions and  thermal properties \citep{Cooper85, Reffet15}.

Section~\ref{background} briefly reviews the relevant aspects of density wave theory, while
Section~\ref{data} provides information about the occultation data used for this investigation. Section~\ref{wavelets} describes the wavelet-based techniques we have developed to isolate and quantify the wave signals associated with particular resonances. Section~\ref{results} discusses the potential wave signatures associated with each of the B-ring Lindblad resonances, and the surface mass densities implied by these features. Section~\ref{discussion} discusses the implications of these new estimates.

\FloatBarrier

\section{Theoretical Background}
\label{background}

Spiral density wave patterns are generated in the rings near Lindblad resonances with periodic gravitational perturbations from either Saturn's various moons or asymmetries in the planet's internal structure. At these locations, the periodic perturbations induce organized radial epicyclic motions in the ring-particles' orbital motions, which in turn generate a spiral wave pattern in the ring's optical depth that propagates through the rings. A good review of the detailed theory behind these structures is provided by  \citet{Shu84}, and we summarize some of the key aspects of these calculations here for the sake of clarity and to introduce the notation used in this paper.

A generic ($m+\ell$):($m-1$) inner Lindblad resonance with a satellite occurs when the ring-particles' radial epicyclic frequency $\kappa$ is an integer multiple of the difference between the frequency of some periodic perturbing force $\Omega_p$ and the particles' mean motion $n$:
\begin{equation}
m(n-\Omega_p)= \kappa
\label{ilrcond}
\end{equation}
In a differentially-rotating disk with finite mass density, these perturbations organize the motions of the ring particles, generating a pattern consisting of $m$ spiral arms that rotates around the planet at the rate $\Omega_p$. 
So long as the fractional optical depth variations $\delta \tau/\tau$ are sufficiently small, they should be described by the following function of ring radius $r$, inertial longitude $\lambda$ and time $t$:
\begin{equation}
\frac{\delta \tau}{\tau}=A(r)\cos\left[\phi_r(r)+\phi_{\lambda t}(\lambda, t)\right],
\label{waveeq}
\end{equation}
where $A(r)$ is a radius-dependent amplitude of the density variations, while $\phi_r$ and $\phi_{\lambda t}$ are phase parameters. Note that $\phi_r$ depends only on the ring radius, and $\phi_{\lambda t}$ depends on a combination of the observed longitude and observation time.  Note that $\phi_r$ governs the overall trends in the pattern's wavelength that are common to all the occultation profiles, while $\phi_{\lambda t}$ affects the exact positions of the wavecrests within each occultation. 


Since the pattern created by the resonance consists of $m$ spiral arms rotating around the planet at the rate $\Omega_p$, the longitude/time-dependent part of the pattern's phase $\phi_{\lambda t}$ can be written as:
\begin{equation}
\phi_{\lambda t}(\lambda, t)=|m|[\lambda-\Omega_p(t-t_0)]
\label{phieq}
 \end{equation}
where $\lambda$ and $t$ are the observed inertial longitude and time, and $t_0$ is a reference epoch time, which for this analysis corresponds to 2008-200T00:00:00 UTC (a time near the middle of the interval spanned by the occultations used in this analysis). For a first-order ($\ell=0$) Lindblad resonance, $\Omega_p$ equals the mean motion of the relevant satellite, but for higher-order resonances the pattern speed is a more complex function of the satellite's orbital parameters. For all Lindblad Resonances, Equation~\ref{ilrcond} can be used  to express the pattern speed in terms of the orbital properties of the ring material:
\begin{equation}
\Omega_p=n(r_L)-\frac{1}{m}\kappa(r_L)=\frac{(m-1)}{m} n(r_L)+\frac{1}{m}\dot{\varpi}(r_L)
\label{omeq}
\end{equation}
where $n(r_L)$, $\kappa(r_L)$ and $\dot{\varpi}(r_L)$ are the orbital mean motion, radial epicyclic frequency and apsidal precession rate at the radial location of the exact resonance $r_L$.  

Meanwhile, the radius-dependent part of the phase $\phi_r$ can be derived from the density-wave's dispersion equation and the resulting radius-dependent radial wavenumber of the pattern $k(r)$. For sufficiently weak waves at sufficiently large distances from the resonance, the perturbations from the density wave should cause the ring's surface mass density to oscillate quasi-sinusoidally as a function of radius with a wavenumber $k(r)$ given by the following expression:
\begin{equation}
k(r) = \frac{d\phi_r}{dr} = \frac{3(m-1)M_P(r-r_L)}{2\pi\sigma_0 r_L^4},
\label{keq}
\end{equation}
where $r_L$ is the radial location of the exact resonance, $M_P$ is the mass of the central planet (Saturn in this case) and $\sigma_0$ is the undisturbed surface mass density of the rings. The explicit dependence on $\sigma_0$ is why measurements of the wave's radial wavenumber at a specified distance from the resonance provide estimates of the rings' local surface mass density. 
Integrating this expression (and assuming a constant mass density), we find that the radius-dependent part of the phase should be given by the following asymptotic expression:
\begin{equation}
\phi_r(r)=\frac{3(m-1)M_P(r-r_L)^2}{4\pi\sigma_0 r_L^4}+\phi_0,
\end{equation} 
where $\phi_0$ is a constant phase offset. 



\section{Observational Data}
\label{data}

\begin{table}
\caption{Longitudes, Ephemeris Time (seconds past J2000 in TDB) and wave phases $\phi_{\lambda t}$ for the occultations by $\gamma$ Crucis used in this analysis (computed using the indicated $m$-numbers and pattern speeds).}
\label{obstab}
\resizebox{5in}{!}{\begin{tabular}{|c||c|c|c|c|c|} \hline Rev & Ja 2:1 & Mi 5:2 & Pd 3:2 & En 3:1 & Ja 3:2 \\ 
& m=2 & m=3 & m=3 & m=2 & m=3 \\
& $\Omega_p=518.24^\circ$/day & $\Omega_p=635.99^\circ$/day & 
$\Omega_p=572.79^\circ$/day & $\Omega_p=393.89^\circ$/day & $\Omega_p=518.24^\circ$/day \\
\hline
&  184.8$^\circ$  &  185.3$^\circ$  &  185.9$^\circ$  &  186.4$^\circ$  &  186.4$^\circ$ \\ 
071 & 266191216. & 266190496.  & 266189456. & 266188496. & 266188384. \\ 
&  142.7$^\circ$  &  156.4$^\circ$  &   55.2$^\circ$  &  303.2$^\circ$  &  269.9$^\circ$ \\ \hline
&  ---   &  ---  &  185.5$^\circ$  &  186.0$^\circ$  &  186.0$^\circ$ \\ 
072 &    ---    &    ---     & 266806000. & 266805040. & 266804928. \\ 
&  ---  &  ---  &   31.7$^\circ$  &   80.7$^\circ$  &  334.2$^\circ$ \\ \hline
&  184.0$^\circ$  &  184.5$^\circ$  &  185.1$^\circ$  &  185.6$^\circ$  &  185.7$^\circ$ \\ 
073 & 267423904. & 267423184.  & 267422144. & 267421184. & 267421088. \\ 
&  113.4$^\circ$  &  292.5$^\circ$  &   16.4$^\circ$  &  222.1$^\circ$  &   45.9$^\circ$ \\ \hline
&  183.0$^\circ$  &  183.6$^\circ$  &  184.2$^\circ$  &  --- &  ---\\ 
077 & 269856064. & 269855328.  & 269854304. &    ---   &    ---   \\ 
&   94.8$^\circ$  &  220.4$^\circ$  &  241.6$^\circ$  &  ---  &  --- \\ \hline
&  182.8$^\circ$  &  183.4$^\circ$  &  184.0$^\circ$  &  184.6$^\circ$  &  184.6$^\circ$ \\ 
078 & 270464544. & 270463808.  & 270462784. & 270461856. & 270461760. \\ 
&  354.9$^\circ$  &  102.7$^\circ$  &   19.2$^\circ$  &  215.8$^\circ$  &   47.9$^\circ$ \\ \hline
&  181.5$^\circ$  &  182.1$^\circ$  &  182.8$^\circ$  &  ---  &  --- \\ 
079 & 271043200. & 271042432.  & 271041344. &    ---   &    ---    \\ 
&  250.2$^\circ$  &  280.9$^\circ$  &   28.9$^\circ$  &  ---  &  --- \\ \hline
&  180.7$^\circ$  &  181.3$^\circ$  &  182.1$^\circ$  &  182.8$^\circ$  &  182.9$^\circ$ \\ 
081 & 272318080. & 272317312.  & 272316224. & 272315232. & 272315104. \\ 
&   75.1$^\circ$  &  205.9$^\circ$  &  231.6$^\circ$  &  233.5$^\circ$  &  172.5$^\circ$ \\ \hline
&  180.3$^\circ$  &  181.0$^\circ$  &  181.8$^\circ$  &  182.5$^\circ$  &  182.5$^\circ$ \\ 
082 & 272953856. & 272953088.  & 272952000. & 272951008. & 272950880. \\ 
&    7.4$^\circ$  &  204.9$^\circ$  &  185.9$^\circ$  &  196.0$^\circ$  &  251.2$^\circ$ \\ \hline
&  179.4$^\circ$  &  180.1$^\circ$  &  181.0$^\circ$  &  181.7$^\circ$  &  181.7$^\circ$ \\ 
086 & 275501376. & 275500640.  & 275499520. & 275498528. & 275498432. \\ 
&   44.8$^\circ$  &  105.2$^\circ$  &  276.9$^\circ$  &    6.4$^\circ$  &  127.5$^\circ$ \\ \hline
&  179.2$^\circ$  &  179.9$^\circ$  &  180.7$^\circ$  &  181.4$^\circ$  &  181.5$^\circ$ \\ 
089 & 277406464. & 277405696.  & 277404608. & 277403616. & 277403488. \\ 
&  230.7$^\circ$  &  154.9$^\circ$  &  187.2$^\circ$  &  275.8$^\circ$  &   46.2$^\circ$ \\ \hline
&  205.9$^\circ$  &  205.2$^\circ$  &  204.4$^\circ$  &  203.8$^\circ$  &  203.7$^\circ$ \\ 
093 & 280042592. & 280041728.  & 280040480. & 280039360. & 280039232. \\ 
&  340.3$^\circ$  &  339.6$^\circ$  &   34.7$^\circ$  &   48.4$^\circ$  &  204.6$^\circ$ \\ \hline
&  191.8$^\circ$  &  191.9$^\circ$  &  191.9$^\circ$  &  192.0$^\circ$  &  192.0$^\circ$ \\ 
094 & 280679008. & 280678208.  & 280677056. & 280676000. & 280675904. \\ 
&  237.6$^\circ$  &  284.0$^\circ$  &  296.6$^\circ$  &  339.8$^\circ$  &  232.9$^\circ$ \\ \hline
&  186.3$^\circ$  &  186.6$^\circ$  &  187.0$^\circ$  &  187.4$^\circ$  &  187.4$^\circ$ \\ 
096 & 282012064. & 282011328.  & 282010272. & 282009312. & 282009216. \\ 
&   75.0$^\circ$  &  349.0$^\circ$  &   46.2$^\circ$  &   53.7$^\circ$  &  346.9$^\circ$ \\ \hline
&  ---  &  186.5$^\circ$  &  186.9$^\circ$  &  187.3$^\circ$  &  187.3$^\circ$ \\ 
097 &    ---    & 282700000.  & 282698944. & 282697984. & 282697888. \\ 
&  ---  &  260.7$^\circ$  &   29.2$^\circ$  &  254.2$^\circ$  &  194.3$^\circ$ \\ \hline
&  218.7$^\circ$  &  217.3$^\circ$  &  215.5$^\circ$  &  214.1$^\circ$  &  213.9$^\circ$ \\ 
100 & 285031424. & 285030592.  & 285029376. & 285028320. & 285028192. \\ 
&  278.7$^\circ$  &    6.8$^\circ$  &  206.0$^\circ$  &  300.5$^\circ$  &  101.9$^\circ$ \\ \hline
&  218.7$^\circ$  &  217.3$^\circ$  &  215.5$^\circ$  &  214.1$^\circ$  &  213.9$^\circ$ \\ 
101 & 285858560. & 285857728.  & 285856544. & 285855456. & 285855328. \\ 
&   76.0$^\circ$  &  100.7$^\circ$  &  315.2$^\circ$  &  318.6$^\circ$  &  337.7$^\circ$ \\ \hline
&  218.4$^\circ$  &  217.0$^\circ$  &  215.3$^\circ$  &  213.8$^\circ$  &  213.7$^\circ$ \\ 
102 & 286683744. & 286682912.  & 286681728. & 286680640. & 286680544. \\ 
&  256.3$^\circ$  &  237.4$^\circ$  &  102.7$^\circ$  &  354.2$^\circ$  &  248.2$^\circ$ \\ \hline \end{tabular}}
\end{table}

\nocite{Acton96}
This analysis uses stellar occultation data obtained by the Visual and Infrared Mapping Spectrometer (VIMS) instrument onboard the Cassini spacecraft \citep{Brown04}. During these observations the instrument measures the brightness of a selected star  repeatedly as it passes behind the rings. While VIMS measures the brightness of the star at multiple wavelengths, for this analysis we use only data obtained at wavelengths around 3 microns, where the ring is especially dark.  Each of these brightness measurements is tagged with a precise time stamp, which (together with the relevant spacecraft trajectory information stored in the NAIF SPICE kernels; Acton 1996) allows us to compute the radius and inertial longitude where the starlight passed through the rings. Based on the positions of sharp edges elsewhere in the rings, we can confirm that these calculations are accurate to within one kilometer. Global fits to the ring geometry provide small corrections to the spacecraft's trajectory that improve this accuracy to within a few hundred meters \citep{French11}. 

This particular investigation examines 17 occultations of the star $\gamma$ Crucis obtained in 2008, corresponding to ``Revs'' (Cassini orbits) 71-102. These occultations are especially useful for studying the B ring because $\gamma$ Crucis is a very bright star that lies in Saturn's far southern skies. The line of sight to the star therefore passes through the rings at a very steep angle ($62.35^\circ$), reducing the light's pathlength through the rings and increasing the signal transmitted through the rings. Furthermore, these occultations were all obtained from very similar observing geometries over a relatively short period of time, which facilitates the comparisons between the various opacity profiles described in Section~\ref{wavelets} below. Table~\ref{obstab} provides a summary of this data set, giving the occultations that cover each of the relevant resonances, along with the inertial longitudes and times where the line of sight to the star crossed the resonant radius $r_L$. 
These numbers, along with the pattern speeds appropriate  for each resonance, are then inserted into Equation~\ref{phieq} in order to compute  the expected $\phi_{\lambda t}$ values for each occultation and each of the waves considered in this study (see Table~\ref{obstab}).

The response of the VIMS instrument is highly linear, so the measured signal is easily translated into estimates of the transmission $T$ through the ring by first subtracting the mean signal in a region where the ring is nearly opaque (105,700-106,100 km from Saturn's center) and then dividing the resulting brightness measurements by the mean signal level in a region unobstructed by ring material (either just outside the B ring in the Huygens Gap  at 117,700-117,750 km or, if these data are missing, outside the entire ring system beyond 145,000 km). This transmission can be converted into estimates of the normal optical depth $\tau_n=-\ln(T)\sin(B)$, where $B=62.35^\circ$ is the elevation angle of the star above the ring plane. Note that the signal from the unocculted star corresponds to about 250 counts  per 100 meters of radius in most of these occultations (with some variability among the occultations due to how well the selected pixel captured the star). Since the read noise of the instrument is low (around 1 count), this means that these occultations have sufficient signal-to-noise to discern sub-percent variations in the transmission on sub-kilometer radius scales, and detect a finite signal through the rings even where the optical depth exceeds 4.

During each occultation, VIMS recorded the average stellar signal every 20-40 ms, which corresponds to a radial range of 200-400 meters. This sampling scale is larger than both the projected stellar diameter (70-100 m) and  the Fresnel zone (60-70 m), and so determines the effective resolution for these observations. In order to simplify comparisons between the profiles and facilitate the multi-profile wavelet analysis described below, the transmission, longitude and time parameters for every profile of each ring feature were re-sampled and interpolated onto a uniform radial grid sampled every 100 meters (well above the resolution of any given occultation). 

\section{Multi-profile wavelet analysis}
\label{wavelets}

The Janus 2:1 and Mimas 5:2 waves are clearly visible in individual occultation profiles, but none of the other density waves can be clearly seen within a single profile. Identifying these waves is difficult not only because the signal levels are low, but also because the relevant parts of the B ring contain intense short-wavelength variations that vary stochastically from occultation to occultation. These variations obscure any coherent signal from the relevant density waves. Fortunately, we can use wavelet-based methods to combine data from multiple profiles and thereby isolate the density wave signals.

The desired density wave patterns have wavelengths that should vary with radius across the ring (see Equation~\ref{keq}), so these waves are most easily identified using wavelets. Continuous wavelet transformations are analogous to localized Fourier transformations and have already proven to be extremely powerful tools for quantifying the properties of waves in planetary rings \citep{Tiscareno07, Colwell09cd, Baillie11, Tiscareno13, HN13, HN14}.  For this investigation, we compute the wavelet transform for each resampled profile with the standard {\tt wavelet} routine in the IDL language \citep{TC98}, using a Morlet mother wavelet with $\omega_0=6$. This yields a wavelet transform $\mathcal{W}_i(r,k)$ for each profile $i$ as a function of radius $r$ and wavenumber $k$.  Note that this is a complex function, and so in general can be written as $\mathcal{W}_i=\mathcal{A}_ie^{i\Phi_i}$, where $\mathcal{A}_i(r,k)$ and $\Phi_i(r,k)$ are the (real) wavelet amplitude and phase. Note that if we have a pure sinusoidal signal at a given wavevector $k_0$, then $\Phi_i(r,k_0)$ is the phase of the wave as a function of position. Also, one can define the wavelet power $\mathcal{P}_i(r,k)=\mathcal{A}_i^2$

\nocite{Foster96}
Previous work by \citet{Colwell09cd} and \citet{Baillie11}  combined wavelet data from multiple occultations in a manner that enabled them to identify waves that were too weak to discern in individual occultation profiles. This method amounts to co-adding the wavelet power from the various occultations (technically, they co-added the values of the WWZ transform, a version of the wavelet power that is better optimized for unevenly sampled data, see Foster 1996). With this approach, peaks in individual wavelet power maps that are due to stochastic opacity variations in individual profiles are averaged out, while persistent signals at particular wavelengths and locations remain in the averaged power. This technique does improve the signal-to-noise for weak waves and is an especially good method for searching for waves with unknown pattern speeds. However, this method does not clearly identify the density wave signals in the central and outer B ring. Fortunately, the waves of interest here have known pattern speeds, enabling us to use a method that is even better at isolating the signatures of density waves.

As mentioned in Section~\ref{background}, spiral density waves are patterns consisting of $m$ tightly wrapped spiral arms rotating around the planet at a predictable pattern speed $\Omega_p$. In a given occultation cut, this pattern produces a quasi-sinusoidal opacity variation described by Equation~\ref{waveeq}. However, depending on the observed longitude and time, each profile will have a different value for the phase parameter $\phi_{\lambda t}$. The crests and troughs of the wave will therefore appear at different locations in different profiles, and the wavelet phases associated with these patterns will vary  from occultation to occultation. Fortunately, the waves considered here are all generated by resonances with known satellites, and so have known values of $m$ and $\Omega_p$, and so we can use Equation~\ref{phieq} to compute the expected $\phi_{\lambda t}$ for each occultation profile and use this information to isolate the desired signal from those specific waves.

Let us denote the calculated value of $\phi_{\lambda t}$ (assuming a given $m$ and $\Omega_p$)  for the $i$-th occultation profile as $\phi_i$. If the track of the star behind the ring was in the radial direction and the star's apparent radial motion across the ring feature were fast compared to the perturbing moon's orbital motion, then $\phi_i$ would have the a constant value for the entire profile. However, in reality both the observation time and observed longitude vary slightly as the star passes behind the wave, so that $\phi_i$ is a function of radius, albeit a very weak one. In any case, we can use $\phi_i$ to compute the {\sl phase-corrected wavelet} for each profile:
\begin{equation}
\mathcal{W}_{\phi,i}(r,k)=\mathcal{W}_i(r,k)e^{-i\phi_i(r)}=\mathcal{A}_i(r,k)e^{i(\Phi_i(r,k)-\phi_i(r))}
\end{equation}
Recall that $\Phi_i$ is the {\sl observed} wavelet phase, while $\phi_i$ is the {\sl expected} longitude/time-dependent part of the wavelet phase for a spiral density wave with the selected $m$-number and pattern speed.  Hence for any signal in the wavelet due to the desired density wave, the corrected phase parameter $\Phi_i-\phi_i$ will equal $\phi_r(r)$  (see Equation~\ref{waveeq}) and should have the same value for all occultations. Thus any signal from such a pattern should persist in the  {\sl average phase-corrected wavelet}:
\begin{equation}
\langle \mathcal{W}_{\phi} (r,k)\rangle=\frac{1}{N}\sum_{i=1}^N\mathcal{W}_{\phi,i}(r,k),
\label{apcw}
\end{equation}
where $N$ is the number of occultations. By contrast, any pattern that does not have the selected $\Omega_p$ should have different phases in the phase-corrected wavelet and thus should average to zero in $\langle \mathcal{W}_{\phi} \rangle$ in the limit of large $N$. The signal-to-noise for the selected waves should therefore be much better in the average phase-corrected wavelet than it is in wavelets from individual profiles. In fact, we find that the average phase-corrected wavelet can yield a clearer detection of weak waves than even the averaged wavelet powers used by \citet{Colwell09cd} and \citet{Baillie11}.

In order the illustrate the utility of the average phase-corrected wavelet, it is useful to consider two distinct wavelet powers. First, we define the {\sl average wavelet power}:
\begin{equation}
\bar\mathcal{P}(r,k)=\langle |\mathcal{W}_{\phi}|^2 \rangle=\frac{1}{N}\sum_{i=1}^N|\mathcal{W}_{\phi, i}|^2.
\end{equation}
This is independent of the individual wavelet phases and so is equivalent to the average value of wavelet powers from the individual profiles $\mathcal{P}_i$:
\begin{equation}
\bar\mathcal{P}(r,k)=\langle |\mathcal{W}_{i}|^2 \rangle=\frac{1}{N}\sum_{i=1}^N|\mathcal{W}_{i}|^2,
\end{equation}
and is therefore similar to the statistics used by \citet{Colwell09cd} and \citet{Baillie11}. Second, define the {\sl power of the average phase-corrected wavelet} as:
\begin{equation}
{\mathcal{P}}_\phi(r,k)=|\langle \mathcal{W}_{\phi}\rangle|^2=\left|\frac{1}{N}\sum_{i=1}^N\mathcal{W}_{\phi,i}\right|^2.
\end{equation}
Note that in this case we perform the averaging prior to taking the absolute square, while the opposite is true for $\bar{\mathcal{P}}$. Recall that for any real variable $x$ the difference $\langle x^2 \rangle-\langle x \rangle^2$ is positive definite and equivalent to the variance of $x$. The difference between these two quantities $\bar{\mathcal{P}}-{\mathcal{P}}_\phi$ is similarly a positive quantity determined by the variance in the real and imaginary components of the wavelet among the various occultations. Hence $\mathcal{P}_\phi$ must have a value between 0 and $\bar{\mathcal{P}}$. In fact, $\mathcal{P}_\phi$ can only equal $\bar{\mathcal{P}}$  when $\Phi_i-\phi_i$ is the same for all occultations, as would be the case if the opacity variations are entirely due to a single wave with the specified pattern speed and $m$-number. For any other signal, $\mathcal{P}_\phi$ will be less than  $\bar{\mathcal{P}}$ because of the finite scatter in $\Phi_i-\phi_i$. Indeed, as $N$ approaches infinity, $\mathcal{P}_\phi$ should approach zero for these signals (provided we can sample all possible values of $\Phi_i-\phi_i$). The ratio between these two powers:
\begin{equation}
\mathcal{R}(r,k)=\frac{\mathcal{P}_\phi(r,k)}{\bar{\mathcal{P}}(r,k)},
\end{equation}
should therefore vary between 1 and 0 depending on how well the phase shifts of the observed data match those derived using Equation~\ref{phieq} for the selected $m$-number and pattern speed. 



\begin{figure}
\centerline{\resizebox{3in}{!}{\includegraphics{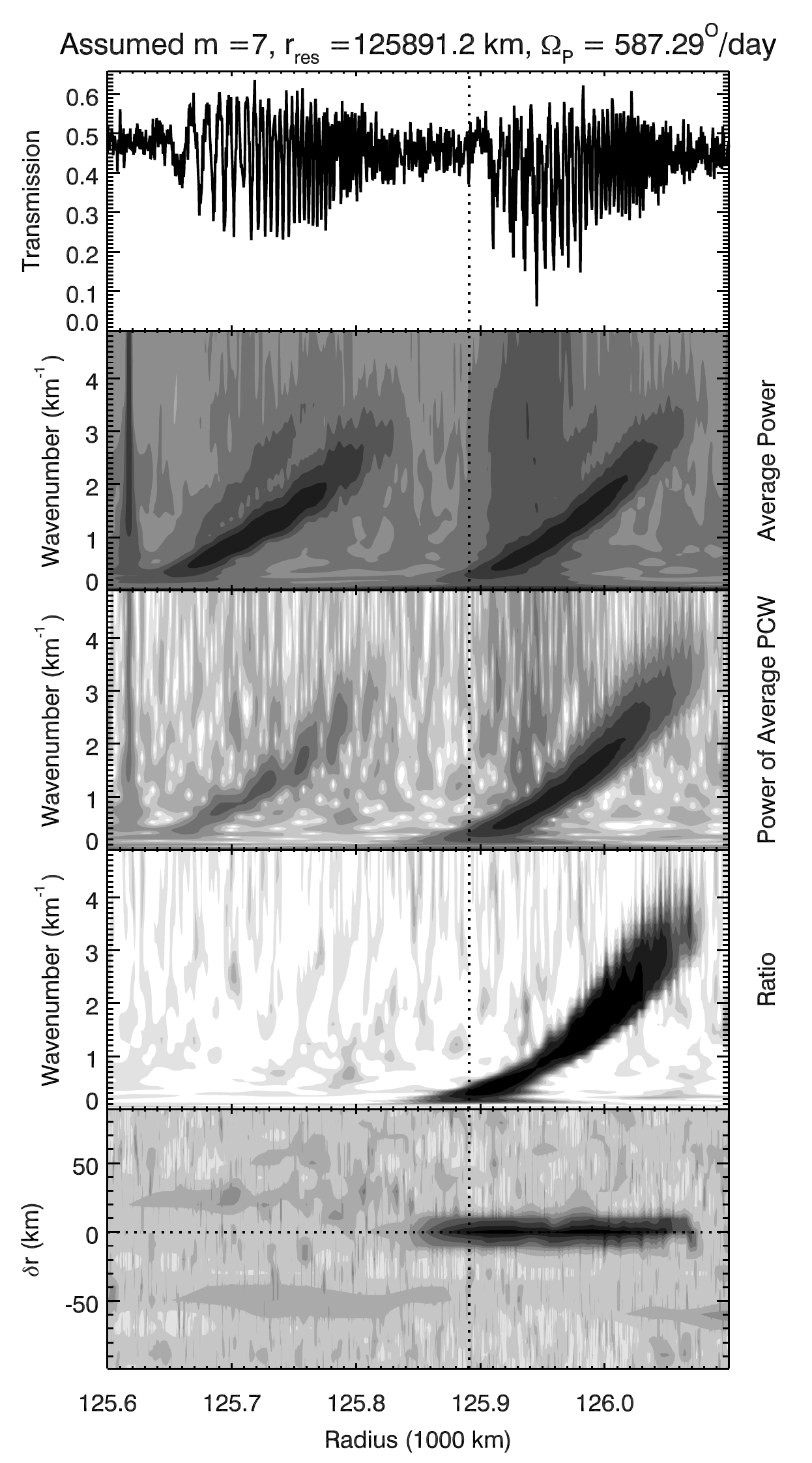}}}
\caption{Sample analysis of the Prometheus 7:6 wave in the A ring. The top panel shows the transmission through the A ring as a function of radius from the Rev 89 occultation by $\gamma$ Crucis. The two density waves clearly visible in this profile are due to the  Pandora 6:5 and Prometheus 7:6 resonances. The second panel shows the average wavelet power $\bar\mathcal{P}$ for the $\gamma$ Crucis occultations, with clear diagonal bands associated with both waves. The third panel shows the power of the average phase-corrected wavelet $\mathcal{P}_\phi$, assuming $m=7$ and a pattern speed appropriate for the Prometheus 7:6 resonance (the exact resonance location is marked by the vertical dotted line). Note that this highlights the right-hand wave. The fourth panel shows the ratio of the above powers $\mathcal{R}$, and shows only the signal from that wave. Finally, the bottom panel shows the peak value of $\mathcal{R}$ as a function of radius and assumed pattern speed, parameterized as a displacement  $\delta r$ from the expected Prometheus 7:6 resonance location (marked with a horizontal dotted line). Note that the maps of $\bar{\mathcal{P}}$ and $\mathcal{P}_\phi$ use a common logarithmic stretch, while the maps of $\mathcal{R}$ use a linear stretch.}
\label{pr76}
\end{figure}

\begin{figure}
\centerline{\resizebox{3in}{!}{\includegraphics{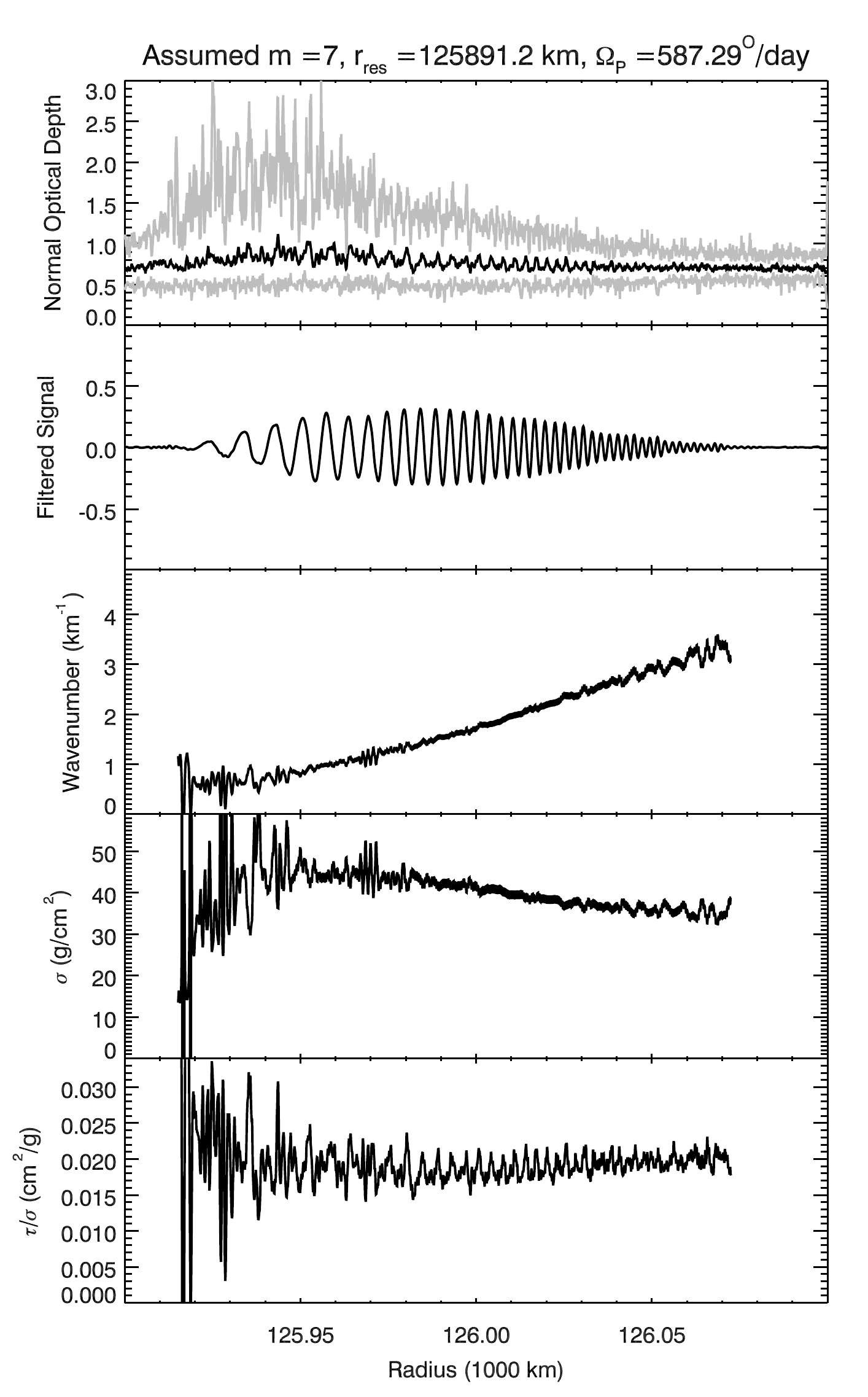}}}
\caption{Extracting wavelength information from the Prometheus 7:6 wave in the A ring from the average phase-corrected wavelet. The top panel shows the mean normal optical depth $\tau_n$ of the ring, along with the range of optical depths among the various profiles. The second panel shows the reconstructed fractional variations in $T$ derived from the average phase-corrected wavelet data for wavelengths between 1 and 10 km. The third panel shows the wavenumber of the pattern as a function of radius. For the sake of clarity, only data where the peak power ratio was above 0.5 are shown. The fourth panel shows the estimated surface mass density $\sigma$ derived from this wave, and the bottom panel shows the estimated opacity $\tau_n/\sigma$. Note that in this case the opacity oscillates because of the residual variations in the average normal optical depth associated with the wave.}
\label{pr76m}
\end{figure}

The utility of these parameters is illustrated in Figure~\ref{pr76}, which shows the results of this sort of analysis for a region of Saturn's A ring occupied by two relatively strong density waves due to the Pandora 6:5 and Prometheus 7:6 resonances. Both of these waves are apparent in the sample profile and can also be seen in the average wavelet power $\bar\mathcal{P}$. However, if we consider the power of the average phase-corrected wavelet $\mathcal{P}_\phi$, then we obtain a different picture. This wavelet power was computed assuming that $m=7$ and $\Omega_p=587.29^\circ$/day, appropriate for the Prometheus 7:6 Lindblad resonance that produces the right-hand wave. As desired, the power of this average phase-corrected wavelet shows a stronger signal from this wave than  it does for the wave generated by Pandora. Furthermore, if we consider the power ratio $\mathcal{R}$, only the signal from the desired Prometheus wave is visible. 

To further validate that the central wave has the expected pattern speed, we can compute the average phase-corrected wavelet for a range of different pattern speeds $\Omega_p$. In practice, we express these pattern speeds in terms of radial displacement $\delta r$ in the assumed resonance location in the rings. For each assumed $\delta r$, we compute the power in the average phase-corrected wavelet and extract the peak power ratio $\mathcal{R}$ at each radius (i.e. the maximum value of $\mathcal{R}$ across all wavenumbers). The bottom panel in Figure~\ref{pr76} displays this peak $\mathcal{R}$ as a function of both radial location in the ring and assumed resonant location. The highest ratios occur along the $\delta r=0$ line within the radial range occupied by the Prometheus 7:6 wave, thus demonstrating that the desired signal only appears in $\mathcal{R}$ when the assumed pattern speed matches the expected value for this wave. 

In addition to isolating signals associated with a particular pattern speed, the average phase-corrected wavelet can be used to quantify the wavelength trends in these signals and thereby obtain estimates of the ring's surface mass density. While in principle the wavelength of the pattern can be determined from the location of the peak power in wavenumber space, in practice the wavenumber can be more robustly determined from the radius-dependent phase of the periodic signal $\phi_r(r)$ \citep{Tiscareno07, HN14o}. In this situation, the best estimator of this phase can be derived
from the average phase-corrected wavelet $\langle{\mathcal W}_\phi(r,k)\rangle$ (Equation~\ref{apcw}). As mentioned above, for any wave signature with the correct pattern speed, the complex phase of this averaged wavelet should equal $\phi_r(r)$. Hence, we may estimate the phase of the wave at a given radius by computing the appropriately weighted average wavelet phase at that location over a range of wavenumbers that encapsulates the desired signal. 

In practice, computing the average phase directly is challenging because the phase parameter is a cyclic quantity. Hence we instead compute the average real and imaginary parts of the wavelet at each radius, and then use these quantities to determine the average phase. More specifically, we 
compute the (complex) profile $D$, whose value at each radius $r$ is given by the following expression:
\begin{equation}
D(r)=\frac{1}{C}\int^{2\pi/k_{max}}_{2\pi/k_{min}}\langle \mathcal{W}_\phi(r,k)\rangle\mathcal{R}(r,k) \frac{d({2\pi}/{k})}{(2\pi/k)^{3/2}}
\end{equation}
where $C$ is a normalization constant. The value of $C$ and the factors of $(2\pi/ k)$ in the integral are chosen so that in the limit where $\mathcal{R}=1$ for all wavenumbers, this expression corresponds to the inverse wavelet transform of $\langle \mathcal{W}_\phi(r,k)\rangle$ \citep{Tiscareno07, TC98}.  Weighting this integral by the power ratio further filters the profiles and thus isolates the desired signal better. This profile $D$ is a complex quantity, with real and imaginary parts $D_R$ and $D_I$ respectively. We therefore define our estimator of the radius-dependent phase of the profile to be $\phi_D(r)=\tan^{-1}(D_I/D_R)$. For a density wave with the relevant pattern speed and $m$-number, $\phi_D(r)=\langle \Phi_i-\phi_i\rangle=\phi_r(r)$. Since $\phi_r(r)$ cycles through $2\pi$ for every cycle of the wave pattern, the radius-dependent wavenumber of the pattern is simply the radial derivative of this phase $k_D(r)=d\phi_D/dr$. Since this estimate of the  wavenumber depends only on the local trends in the phase parameter, it can be used to estimate the ring's local surface mass density via Equation~\ref{keq}.

Figure~\ref{pr76m} illustrates these procedures for the Proemtheus 7:6 wave in the A ring. The top panel of this figure shows the maximum, minimum and mean normal optical depths derived from the relevant occultation profiles. The second panel shows the real part of the $D$ profile derived from the average phase-corrected wavelet. For this particular profile, we integrated the wavelet between wavenumbers of $2\pi/1$ km and $2\pi/10$ km.  The plotted quantity is actually $D_R/\bar{T}$, where $D_R$ is the real part of the profile computed from the average phase-corrected wavelet, and $\bar{T}$ is the average transmission through the ring among all the profiles. This plot therefore shows the fractional variations in the transmission associated with the wave. Note that while the signals from the other waves are outside the range of this particular plot, they would be strongly attenuated in this filtered profile. The third panel plots the pattern's wavenumber derived from the real and imaginary parts of this profile. These data are only plotted where the peak power ratio $\mathcal{R}$ exceeds 0.5 for the sake of clarity. The trend of increasing wavenumber with distance from the resonance is clear. Finally the bottom two panels show estimates of the surface mass density $\sigma$ and the opacity $\tau_n/\sigma$ derived from these wavenumber estimates. There is some scatter in these values (especially where the wavenumber is small close to the resonance). Nevertheless, these data indicate that the surface mass density is between 35 g/cm$^2$ and 50 g/cm$^2$, which is reasonably consistent with the values of $42.4\pm 0.2 $ g/cm$^2$  and $45.1\pm12.2$ g/cm$^2$ that \citet{Spilker04} derived using Voyager PPS and RRS occultation data. Our numbers are also similar to the mass densities derived from weaker nearby waves by \citet{Tiscareno07}.

\section{Results}
\label{results}

The above procedures were used to search for density waves associated with the six strongest Lindblad resonances in Saturn's B ring: the Janus 2:1 resonance at 96,247 km, the Mimas 5:2 resonance at 101,311 km, the Prometheus 3:2 resonance at 106,772 km,  the Pandora 3:2 resonance at 108,546 km, the Enceladus 3:1 resonance at 115,207 km, and the  Janus 3:2 resonance at 115,959 km. We could not identify any potential wave signature associated with the Prometheus 3:2 resonance, but did find interesting signals for all of the other waves. Each of these patterns is discussed below, beginning with the Mimas 5:2 wave, which provides the cleanest and simplest density wave signal anywhere in the B ring. Then we consider the Janus 2:1 wave, whose pattern speed shows surprising irregularities that are likely tied to the periodic changes in Janus' mean motion. Finally, we discuss the evidence for the Janus 3:2, Enceladus 3:1 and Pandora 3:2 patterns, which may represent previously undetected density waves in regions of extremely high optical depth.

\subsection{The Mimas 5:2 wave}

\begin{figure}
\centerline{\resizebox{3in}{!}{\includegraphics{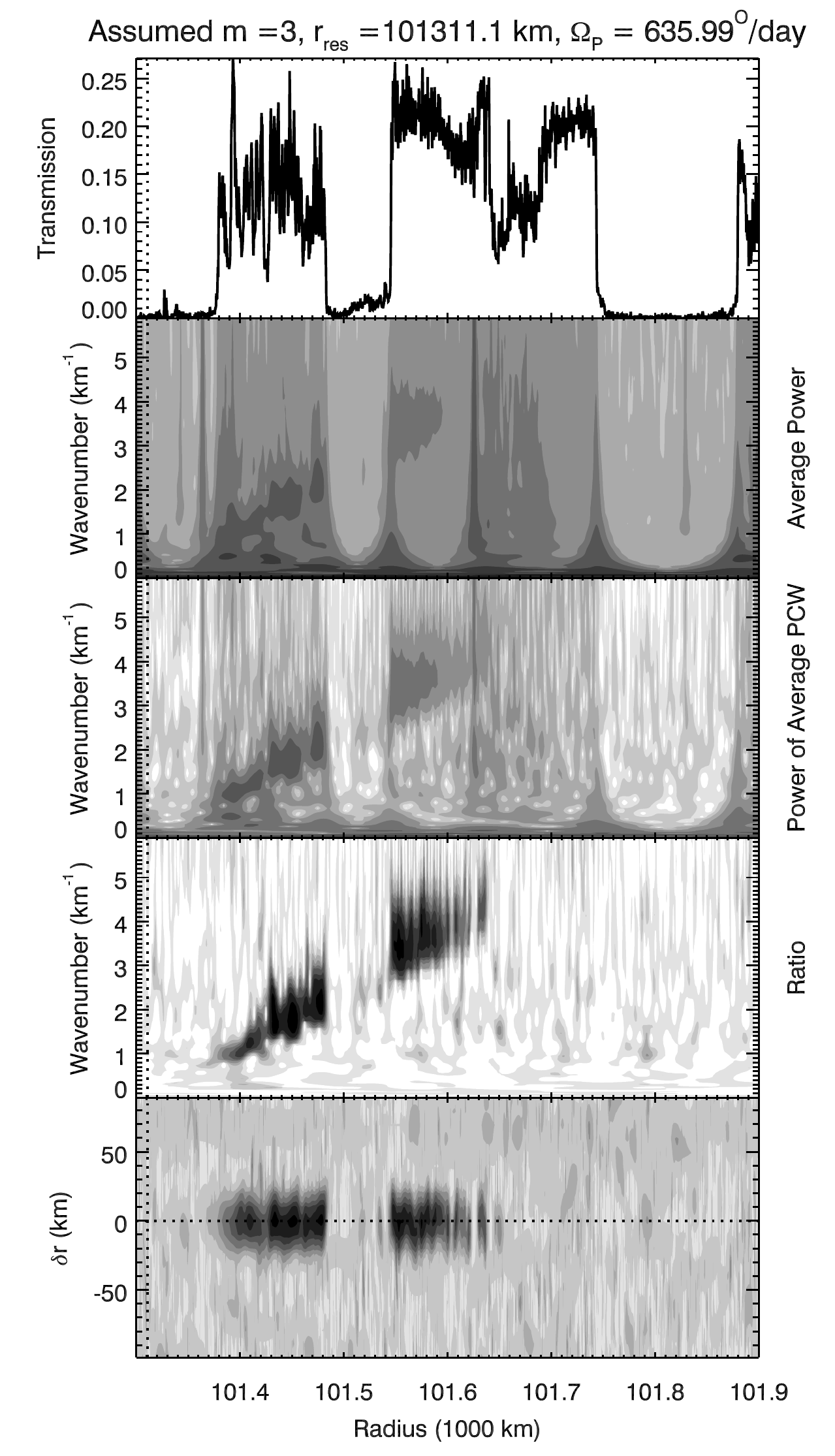}}}
\caption {
Wavelet analysis of the Mimas 5:2 wave in the same format as Figure~\ref{pr76}. The signature of the Mimas 5:2 wave can be seen in the middle three panels as a diagonal dark band extending across the region occupied by the two innermost regions of reduced optical depth out to about 101,650 km. Note also that the signal is strongest at the expected pattern speed for this resonance (i.e $\delta r=0$) in the bottom panel.} 
\label{mi52wave}
\end{figure}

\begin{figure}
\centerline{\resizebox{3in}{!}{\includegraphics{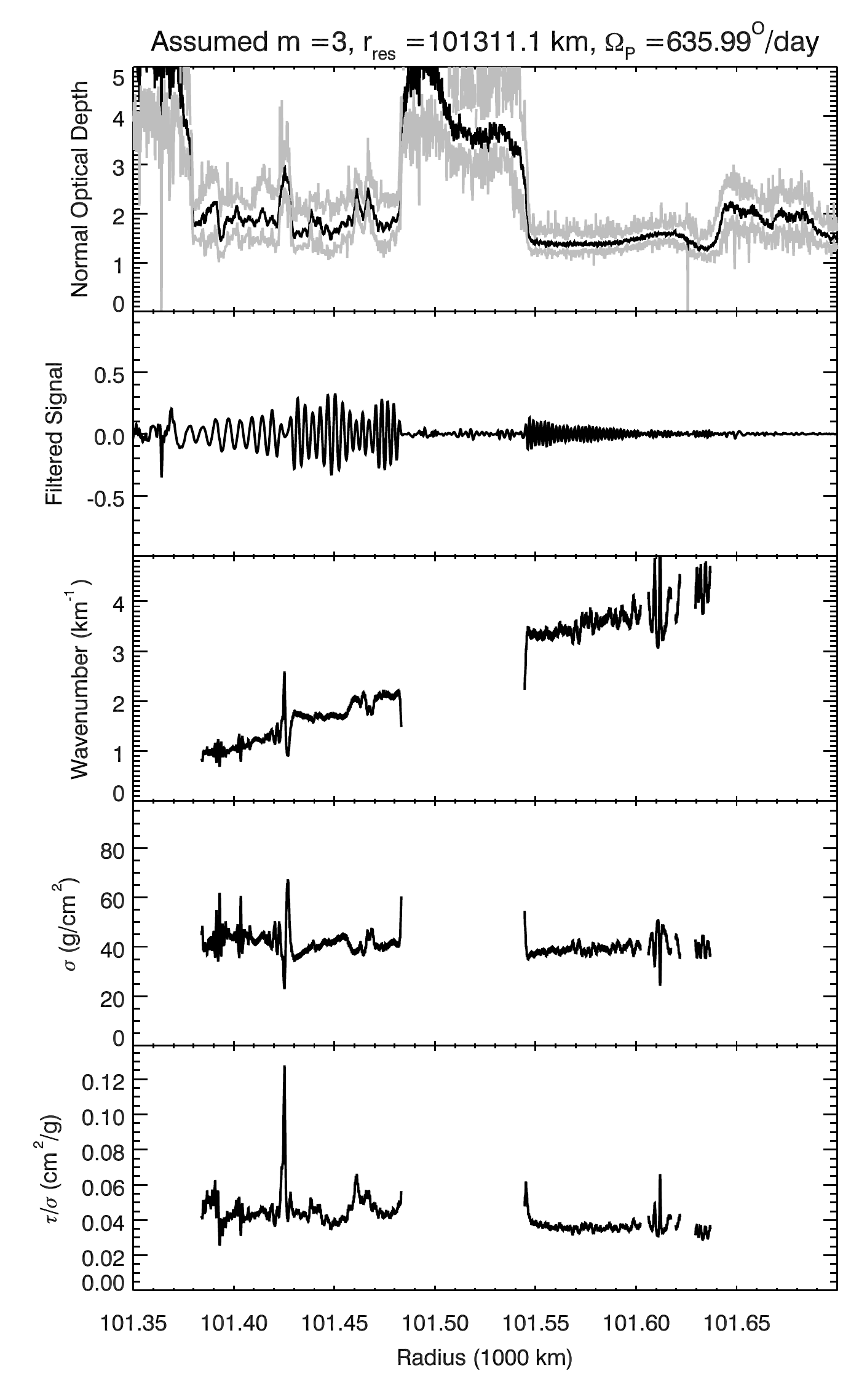}}}
\caption{Parameters for the Mimas 5:2 wave derived from the average phase-corrected wavelet in the same format as Figure~\ref{pr76m}. In this case the reconstructed profile is derived from the average phase-corrected wavelet data for wavelengths between 1 and 10 km, and only data where the peak power ratio was above 0.5 are shown in the bottom three panels.  Note that the resonant radius is off the left edge of the plot, in an opaque region}
\label{mi52m}
\end{figure}

The Mimas 5:2 resonance falls at 101,311 km in the BII region, where the ring rapidly and repeatedly shifts between a state with an optical depth around 2 and one that is nearly opaque. The 5:2 resonance with Mimas falls in one of the opaque regions, but the wave launched by this resonance can be clearly  seen propagating  across two regions with optical depth around 2 between 101,370 km and 101,650 km.  Figure~\ref{mi52wave} shows the wavelet powers and power ratios derived from an analysis of this wave (in the same format as Figure~\ref{pr76}) assuming $m=3$ and a pattern speed of 635.99$^\circ$/day, which is appropriate for this resonance.\footnote{Note that this is a third order resonance and so the expected pattern speed is $\Omega _p=(5n_M-2\dot{\varpi}_M)/3$, where $n_M$ and $\dot{\varpi}_M$ are Mimas' mean motion and apsidal precession rate, respectively.} The wave can clearly be seen as a diagonal streak in the average wavelet power, but this signature is somewhat contaminated by signals associated with other optical depth structures. In particular, the rapid optical depth transitions produce power over a broad range of wavelengths, obscuring the wave signal. By comparison, the wave signal is much clearer in the power of the average phase-corrected wavelet because the phase corrections add a range of phases to the signals from fixed features like sharp edges, and so these signals partially cancel out when the wavelets are averaged together. The ratio of the two wavelet powers provides a even cleaner picture of the wave signature in both regions with finite transmission. A faint hint of a signal can also be seen between 101,520 and 101,550 km, but closer inspection of these data did not show any convincing evidence of a coherent wave signal in this region (although some profiles show one or more wave-like peaks). Finally, the bottom panel shows that these strong ratios only occur for pattern speeds close to the expected value for the 5:2 resonance.

Figure~\ref{mi52m} shows the wave profile derived from the average phase-corrected wavelet (computed by averaging over wavenumbers between $2\pi/1$ km and $2\pi/10$ km), along with the estimates of the wavenumber, surface mass density and $\tau_n/\sigma$ as functions of radius (only data where the peak power ratio was above 0.5 are plotted for the sake of clarity).  Both regions where the wave is clearly visible exhibit surface mass densities close to 40 g/cm$^2$ (see also Table~\ref{masstab}). This number might at first seem surprisingly low, since it is comparable to typical values for the A ring \citep{Tiscareno07}, even though the wave occupies regions with  optical depths 3-4 times higher than those of the A ring. However, this number is not much different from the $54\pm10$ g/cm$^2$ \citet{Lissauer85} derived from the Mimas 4:2 bending wave, which lies in a region of comparable optical depth.

\subsection{Janus 2:1 Wave}

\begin{figure}
\centerline{\resizebox{3in}{!}{\includegraphics{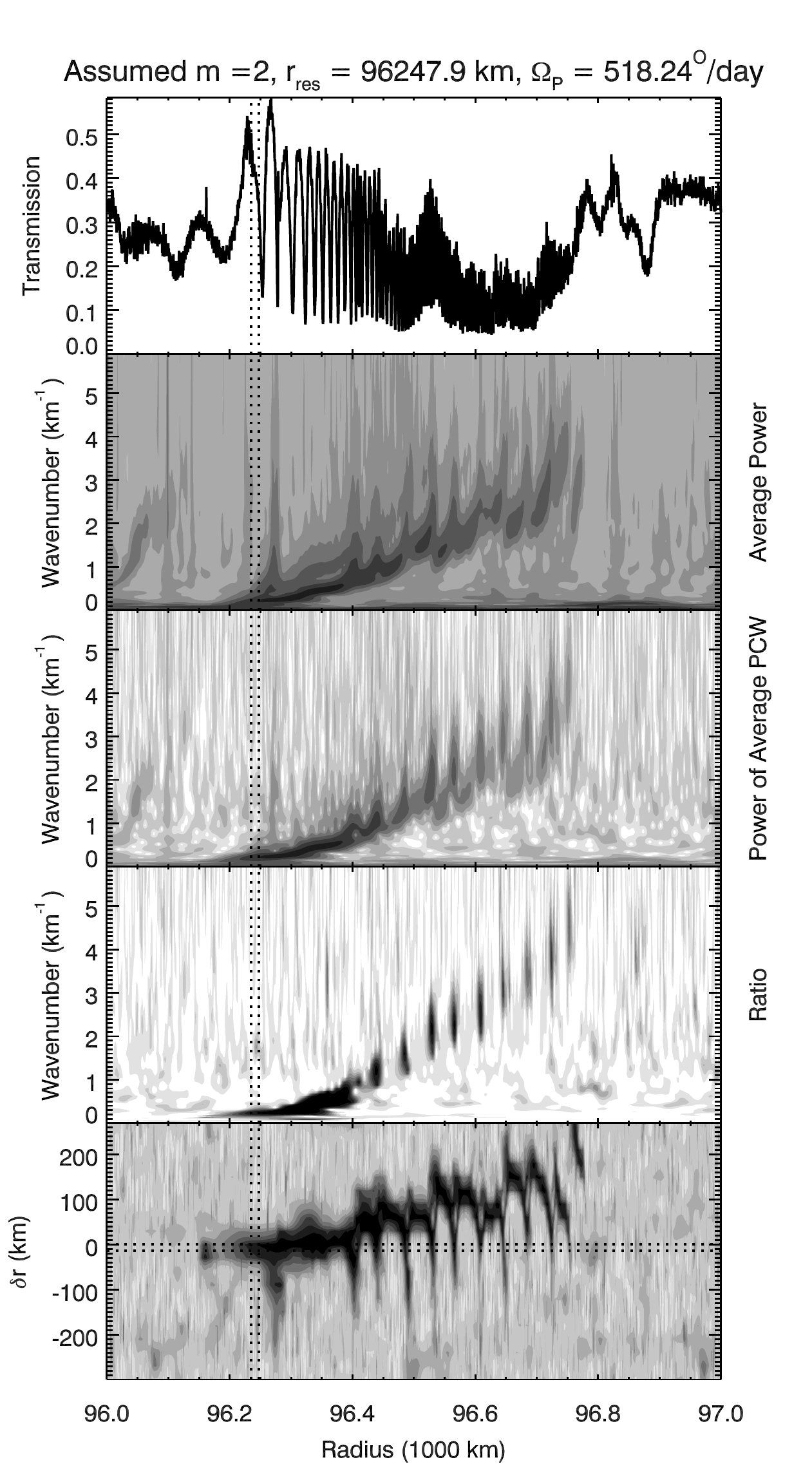}}}
\caption {Wavelet analysis of the Janus 2:1 wave in the same format as Figure~\ref{pr76}. 
The signature of the Janus 2:1 wave is visible in all three wavelet plots, but in the lower two there are ``gaps" in the wave signature. These regions correspond to parts of the wave that have pattern speeds that deviate from the expected value. The two horizontal dotted lines in the bottom panel correspond to the two different resonant radii. Note that the apparent resonance location of this wave shifts outwards as the wave propagates.}
\label{ja21wave}
\end{figure}
 
The Janus 2:1 wave is the most obvious density wave in the B ring, and has already been used to estimate the mass density of the BI region \citep{Holberg82, Esposito83}. Thus it is not surprising that the wave signature is very clear in the average wavelet power, as shown in Figure~\ref{ja21wave}. However, unlike the Mimas 5:2 and Prometheus 7:6 waves considered above, the signature of the Janus 2:1 wave in the average wavelet power is not a continuous diagonal band, but instead a more jagged pattern.  This more complex wave structure almost certainly arises from Janus' unusual orbital properties. Every four years, the semi-major axis and mean motion of Janus alternates between one of two values due to its gravitational interactions with its co-orbital companion Epimetheus \citep{Yoder83}. These shifts in the moon's position cause the location of the 2:1 resonance  to oscillate between 96,235 and 96,248 km, and the relevant pattern speed switches between 518.35$^\circ$/day and 518.24$^\circ$/day. During the time of the observations, Janus was interior to Epimetheus and so the faster pattern speed was active. However, since the wave propagates away from the resonance at a finite speed, some parts of the wave were generated when Janus was exterior to Epimetheus. Interference between these different wave segments with different pattern speeds are thought to be responsible for abrupt ``glitches'' seen in the wave profile \citep{Porco05, Tiscareno06}. 

Given that the nominal resonance location swaps between two discrete locations, one
might reasonably expect that the pattern speed of the wave would oscillate between two different values. However, the real situation turns out to be quite different. Figure~\ref{ja21wave} shows the power of the average phase-corrected wavelet and the power ratio assuming a pattern speed of 518.24$^\circ$, which corresponds to the expected pattern speed of the Janus 2:1 resonance when Janus is outside of Epimetheus' orbit. The signature of the wave still extends across the same range of radii, but is discontinuous, indicating that some parts of the wave have a different pattern speed. Surprisingly,  these other parts of the wave do {\em not} have pattern speeds consistent with the faster rate one would predict for a pattern generated by a resonance with Janus in its second configuration (interior to Epimetheus' orbit). Instead, the pattern speed of the other parts of this wave appear to be {\em slower} than Janus' mean motion in either configuration. This is most clearly illustrated by the bottom panel of Figure~\ref{ja21wave}, which reveals that  the nominal resonance location of the peak power ratio tends to move outwards with increasing radius, indicating a general decrease in the wave's pattern speed. Furthermore, these changes in pattern speed appear to be discontinuous, with abrupt shifts of order 50 km in $\delta r$ (corresponding to 0.4$^\circ$/day in $\Omega_p$) separated by narrow regions where the pattern speed returns to its nominal value. 

\begin{figure}
\centerline{\resizebox{4in}{!}{\includegraphics{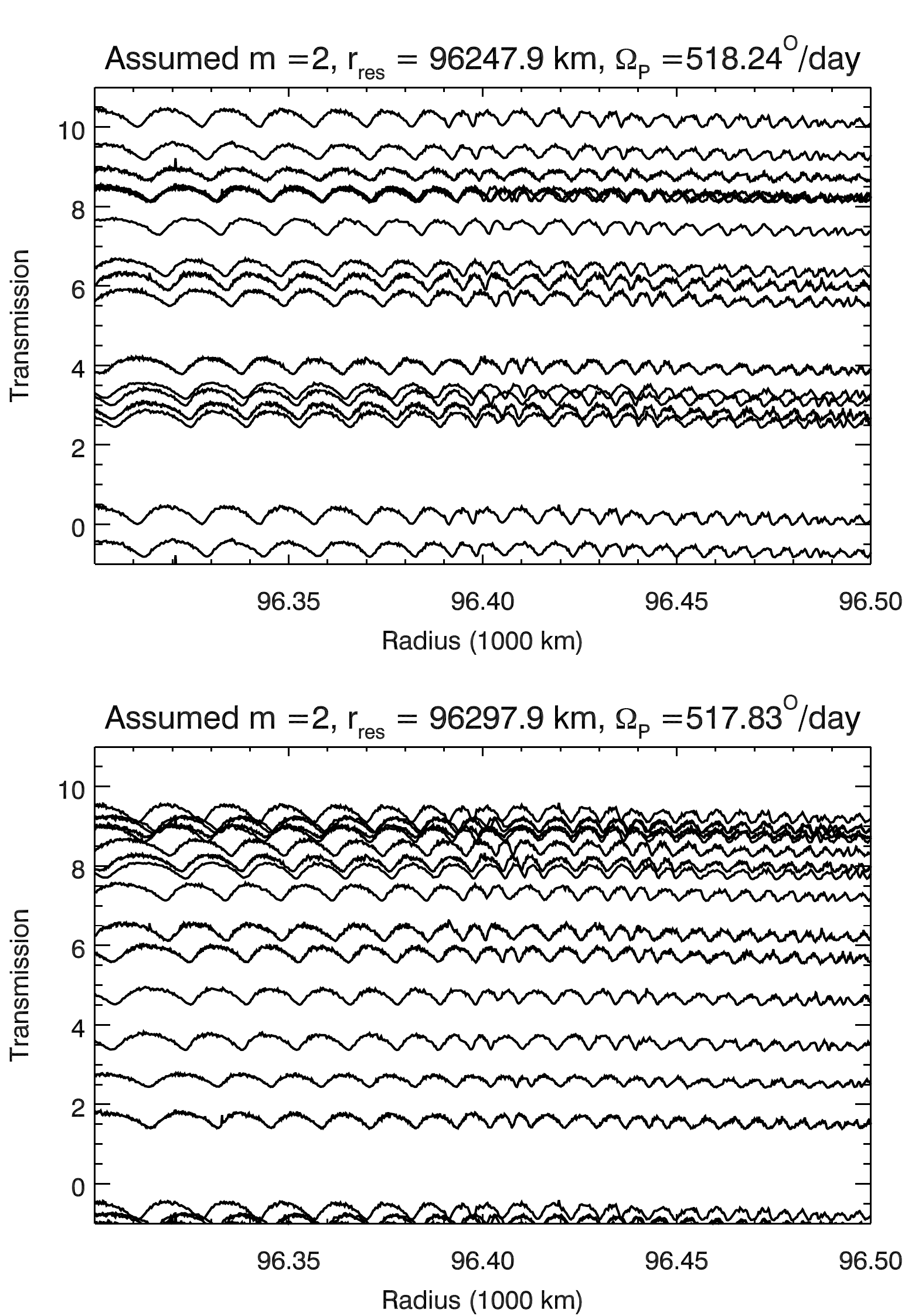}}}
\caption{Different parts of the Janus 2:1 wave have different pattern speeds. In both the above plots, the occultation profiles are shown with vertical offsets proportional to the phase parameter $\phi_{\lambda t}$, computed assuming a given $m$ and pattern speed. In the top panel the pattern speed is near the expected value for the Janus 2:1 resonance, but for the bottom one the pattern speed corresponds to  a radius 50 km further from the planet. For radii less than 96,400 km, the nominal pattern speed better organizes the data, while for the region exterior to 96,400 km the slower pattern speed does a better job, consistent with the data shown in Figure~\ref{ja21wave}. Note the transition between these two regions contains an extra narrow dip.}
\label{ja21profs}
\end{figure}

An examination of the raw profiles confirms these changes in the wave's pattern speed. Figure~\ref{ja21profs} shows the relevant occultation profiles, offset vertically be an amount proportional to the phase parameter $\phi_{\lambda t}$ with two different assumed pattern speeds (note that no phase shifts have been applied to these data). If the pattern speed properly organized the data,  profiles with similar phase shifts (and similar vertical offsets) should be aligned with each other, and the positions of peaks and troughs should shift systematically as the phase changes. In the upper panel of this plot, where the pattern speed matches that expected for the Janus 2:1 resonance, we find the profiles are reasonably well organized for radii less than 96,400 km, but beyond this point there are several profiles which should be close in phase but have peaks and troughs in very different locations. Conversely, if we choose a slower pattern speed, the profiles exterior to 96,400 km line up well but the data interior to 96,400 km show inconsistencies. Note that around 96,400 km, there is a ``glitch" in the wave profile, where two dips occur closer to each other than they do just inside or outside this region. This probably represents the edge of one of the coherent wave segments generated during a time when Janus was at one particular semi-major axis \citep{Tiscareno06}. 

We do not yet have a complete explanation for how different parts of this wave can have such different pattern speeds, but this strange behavior is almost certainly the result of Janus' periodic orbit changes. When Janus' orbit changes, the  radial location where the wave is generated suddenly moves, giving rise to ``glitches'' in the wave profile that propagate slowly outward at the group velocity  $v_g=\pi G\sigma/\kappa$
\citep{Shu84, Tiscareno06}. These glitches interrupt the spiral wave pattern and thus disrupt the gravitational interactions between different parts of the wave that allow it to form a  coherent spiral pattern with a common pattern speed, and so  might allow the distal portions of the wave to become partially decoupled from the resonance. In this situation, the pattern speed of the wave might shift away from $n(r_L)-\kappa(r_L)/m$ and towards $n(r)-\kappa(r)/m$, where $n(r)$ an $\kappa(r)$ are the local mean motion and epicyclic frequency. However, the decoupling from the resonance cannot be complete, since the pattern speeds never get as slow as $n(r)-\kappa(r)/m$, which would correspond to $\delta r =r-r_L$ in the bottom panel in Figure~\ref{ja21wave}. Furthermore, even though the wave exhibits visible glitches every 50 km or so, not every glitch leads to a systematic change in the wave's pattern speed (for example, the glitches at around 96,440 km in Figure~\ref{ja21profs} do not appear to separate two regions of very different pattern speeds). Clearly, much more work will be needed before the dynamics of this unusual wave can be fully understood.

\begin{figure}
\centerline{\resizebox{3in}{!}{\includegraphics{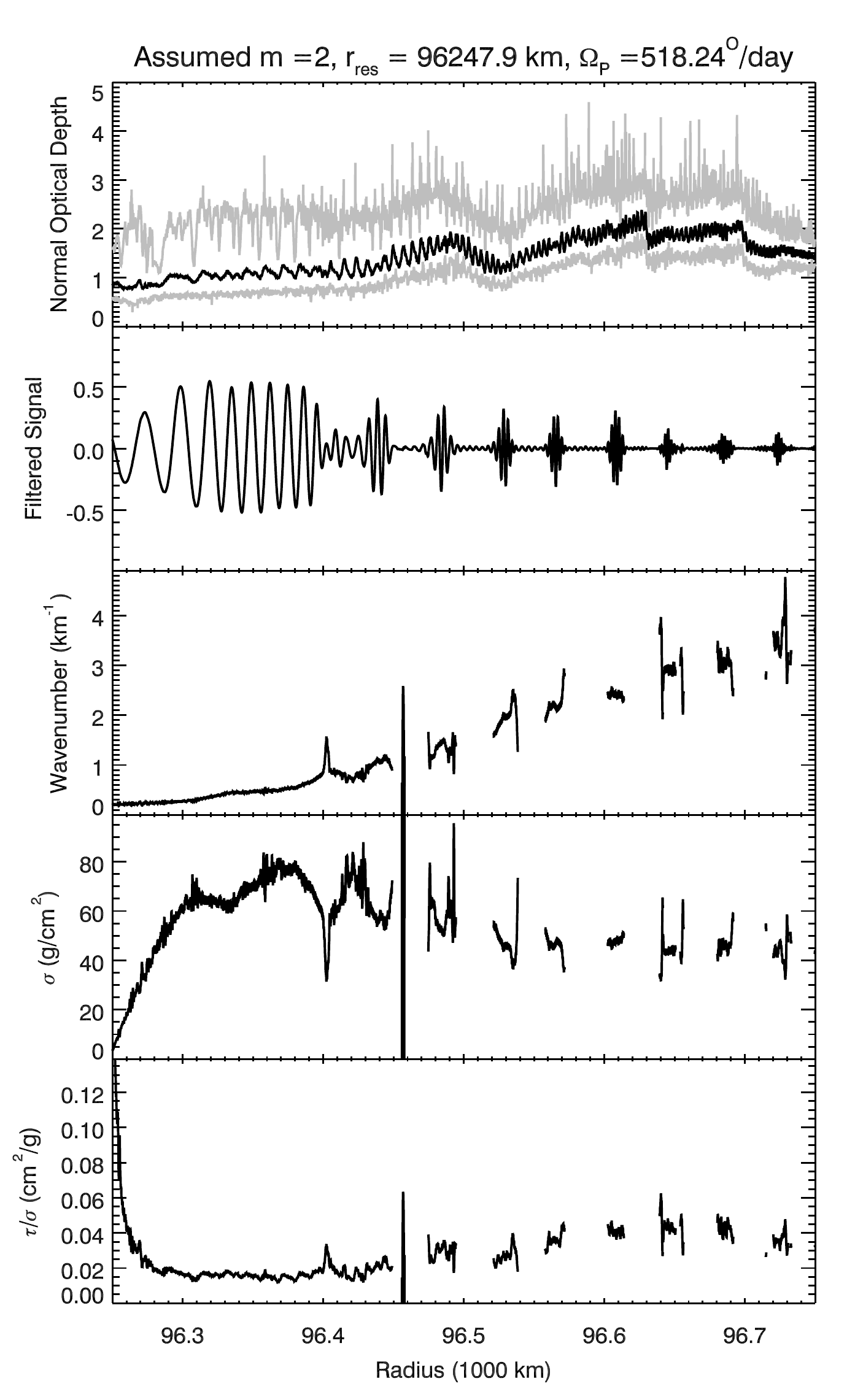}}}
\caption{Parameters for the Janus 2:1 wave derived from the average phase-corrected wavelet assuming a pattern speed close to the expected value for the Janus 2:1 resonance, in the same format as Figure~\ref{pr76m}.  In this case the reconstructed profile is derived from the average phase-corrected wavelet data for wavelengths between 1 and 100 km, and only data where the peak power ratio was above 0.3 are shown in the bottom three panels.}
\label{ja21ma}
\end{figure}

\begin{figure}
\centerline{\resizebox{3in}{!}{\includegraphics{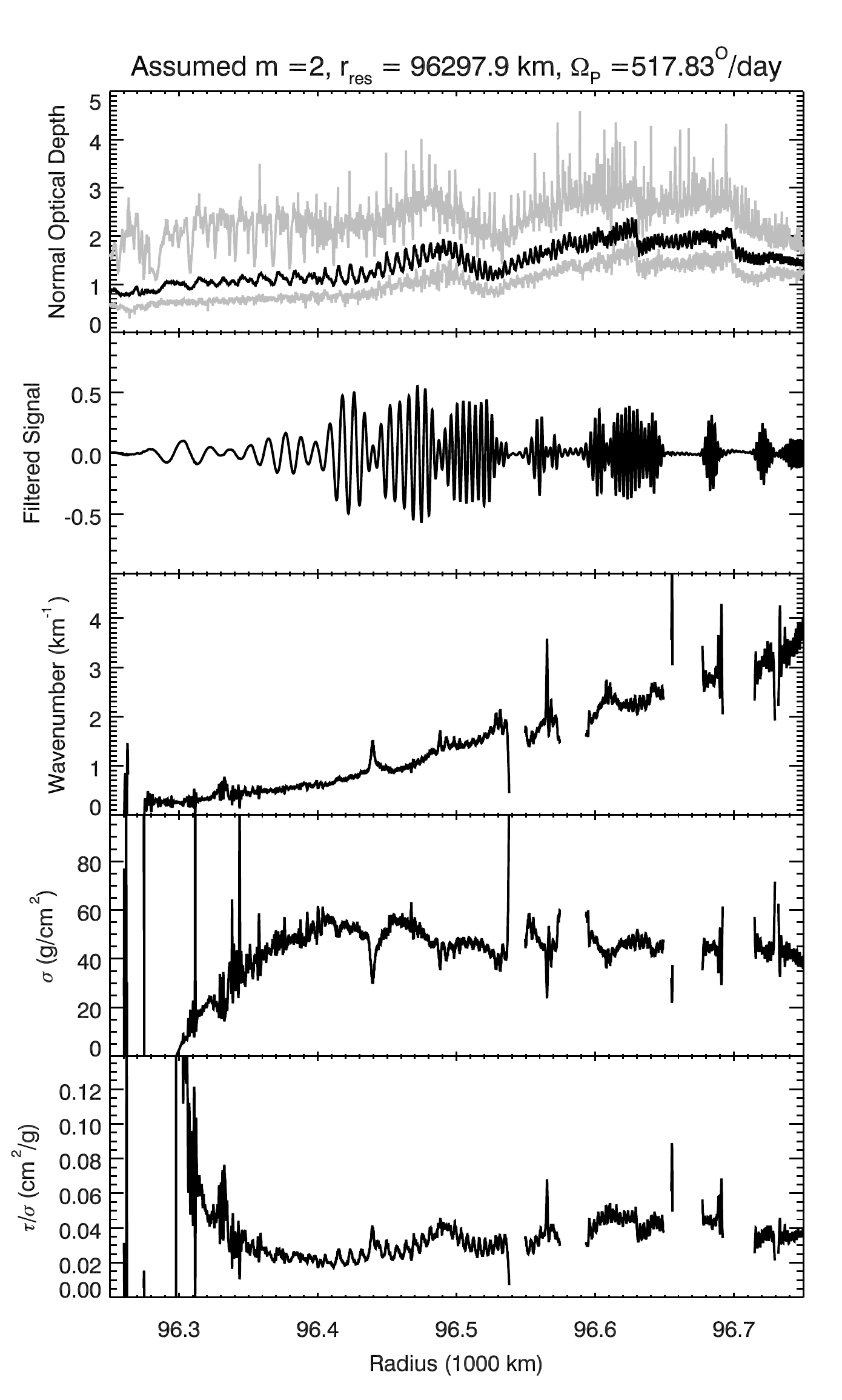}}}
\caption{Parameters for the Janus 2:1 wave derived from the average phase-corrected wavelet assuming a pattern speed 0.4$^\circ$/day slower than that predicted for the Janus 2:1 resonance, in the same format as Figure~\ref{pr76m}.  In this case the reconstructed profile is derived from the average phase-corrected wavelet data for wavelengths between 1 and 100 km, and only data where the peak power ratio was above 0.3 are shown in the bottom three panels.}
\label{ja21mb}
\end{figure}

Given that this wave is not propagating exactly like a normal density wave, the mass estimates derived from this feature might be questionable. However, this is the only density wave in the B ring that has yielded published mass density estimates, so analyzing this wave is still useful, if only for verifying our algorithms. Figures~\ref{ja21ma} and~\ref{ja21mb} show the reconstructed wave and derived parameters for two different assumed pattern speeds, one appropriate for the resonance and one 0.4$^\circ$/day slower (i.e. the same pattern speed showed in the lower panel of Figure~\ref{ja21profs}). These two pattern speeds yield quite different reconstructed profiles, as is to be expected given that different parts of the wave will be better organized by one of these two options than by the other. Interestingly, however, the pattern's wavenumber, surface mass density and $\tau_n/\sigma$ follow very similar trends exterior to 96,400 km (interior to 96,400 km, the two curves diverge, likely because the slower pattern speed places the resonance within the wave, and so Equation~\ref{keq} is not appropriate). This suggests that the mass density estimates are not very sensitive to the assumed pattern speed. Furthermore, when we assume the predicted pattern speed for the resonance, the surface mass density peaks at around 70 g/cm$^2$, consistent with previous estimates \citep{Holberg82, Esposito83}. However, for either pattern speed, the mass density  appears to drop to around 40 g/cm$^2$ further from the resonance. A mass density of 40-70 g/cm$^2$ yields a group velocity of 13-22 km/year, so during each four-year period when Janus has a nearly constant mean motion, it should produce a wave segment between 50 and 90 km wide. If we also account for the 13-km shifts in the location of the resonances, this is compatible with the observed 50-100 km widths of the regions between the glitches in the wave profile, and so the mass density estimates derived from the wavelet analysis do not appear to be wildly off.

\subsection{Previously unidentified waves}

\begin{figure}
\centerline{\resizebox{3in}{!}{\includegraphics{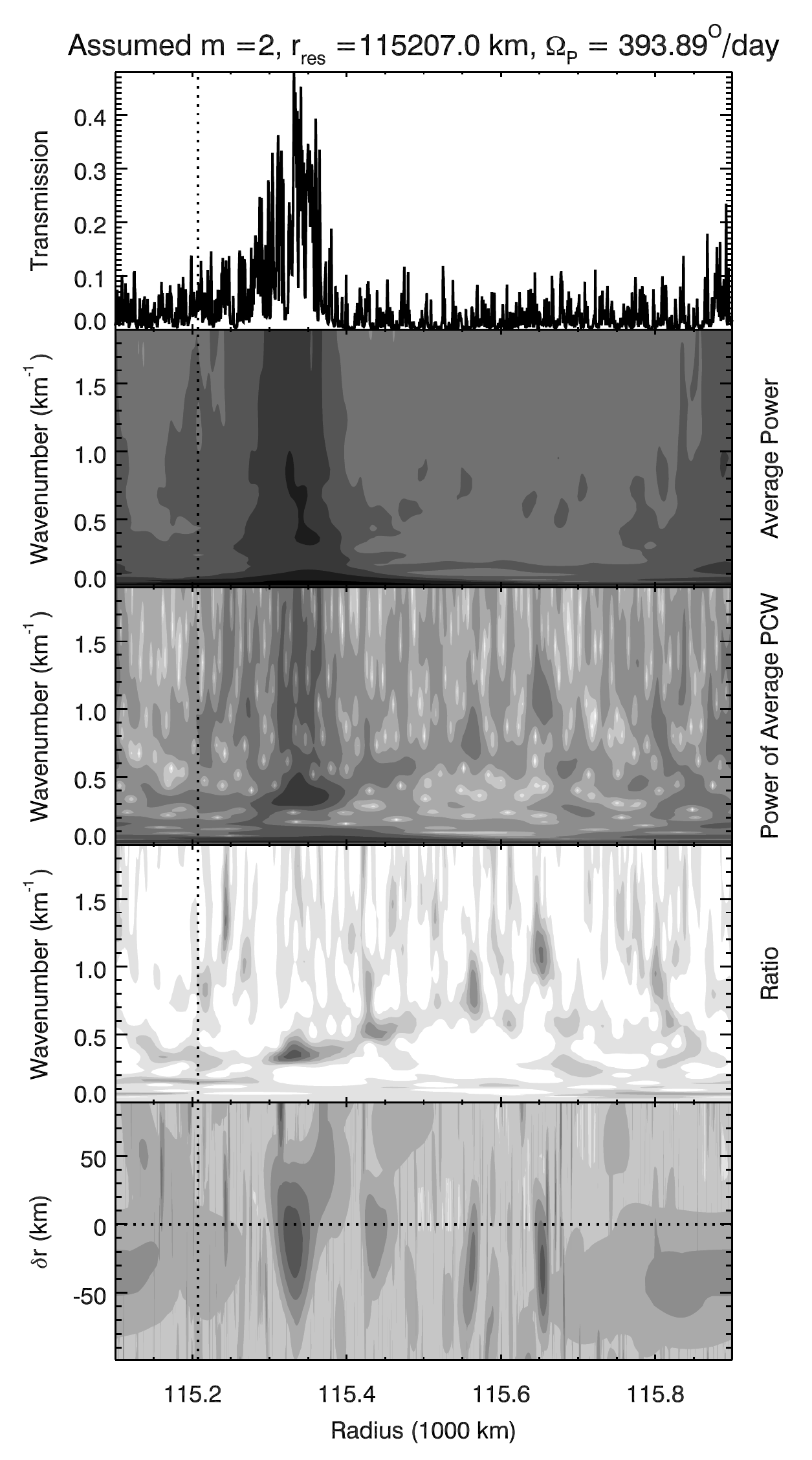}}}
\caption {Wavelet analysis on the Enceladus 3:1 wave in the same format as Figure~\ref{pr76}. 
A weak wave-like signature can be observed between 115,300 and 115,650 km in the ratio plot. The bottom panel demonstrates that this signal only occurs when the assumed pattern speed is fairly close to the expected pattern speed of the density wave (i.e. where $\delta r$ is close to zero).}
\label{en31wave}
\end{figure}

\begin{figure}
\centerline{\resizebox{3in}{!}{\includegraphics{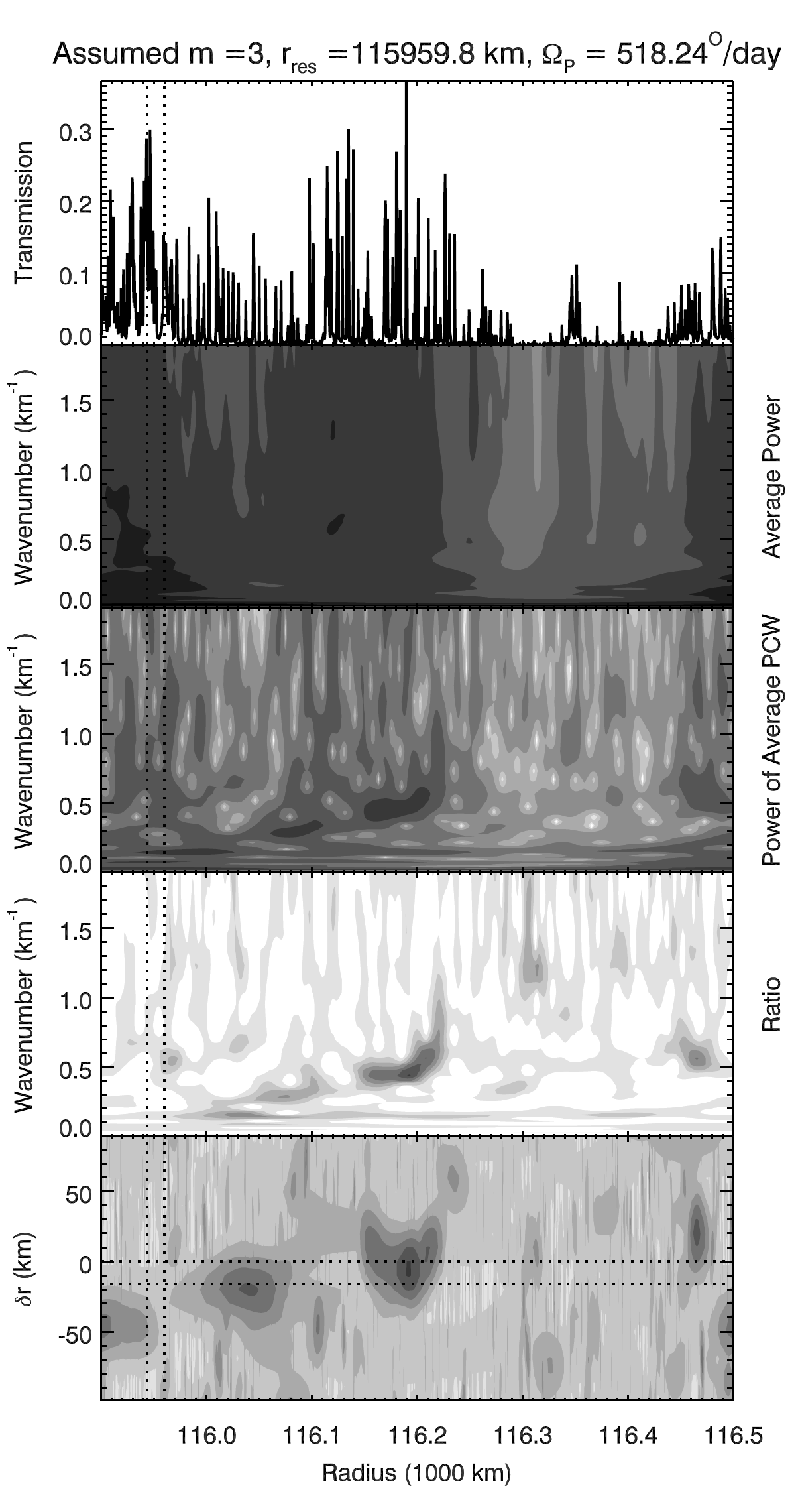}}}
\caption {Wavelet analysis on the Janus 3:2 wave in the same format as Figure~\ref{pr76}.
A weak wave-like signature can be observed between 116,030 and 116,220 km in the ratio plot. This signal is strongest where $\delta r=0$, i.e. when the assumed pattern speed is close to the expected pattern speeds for this density wave (shown as two horizontal dotted lines).}
\label{ja32wave}
\end{figure}

\begin{figure}
\centerline{\resizebox{3in}{!}{\includegraphics{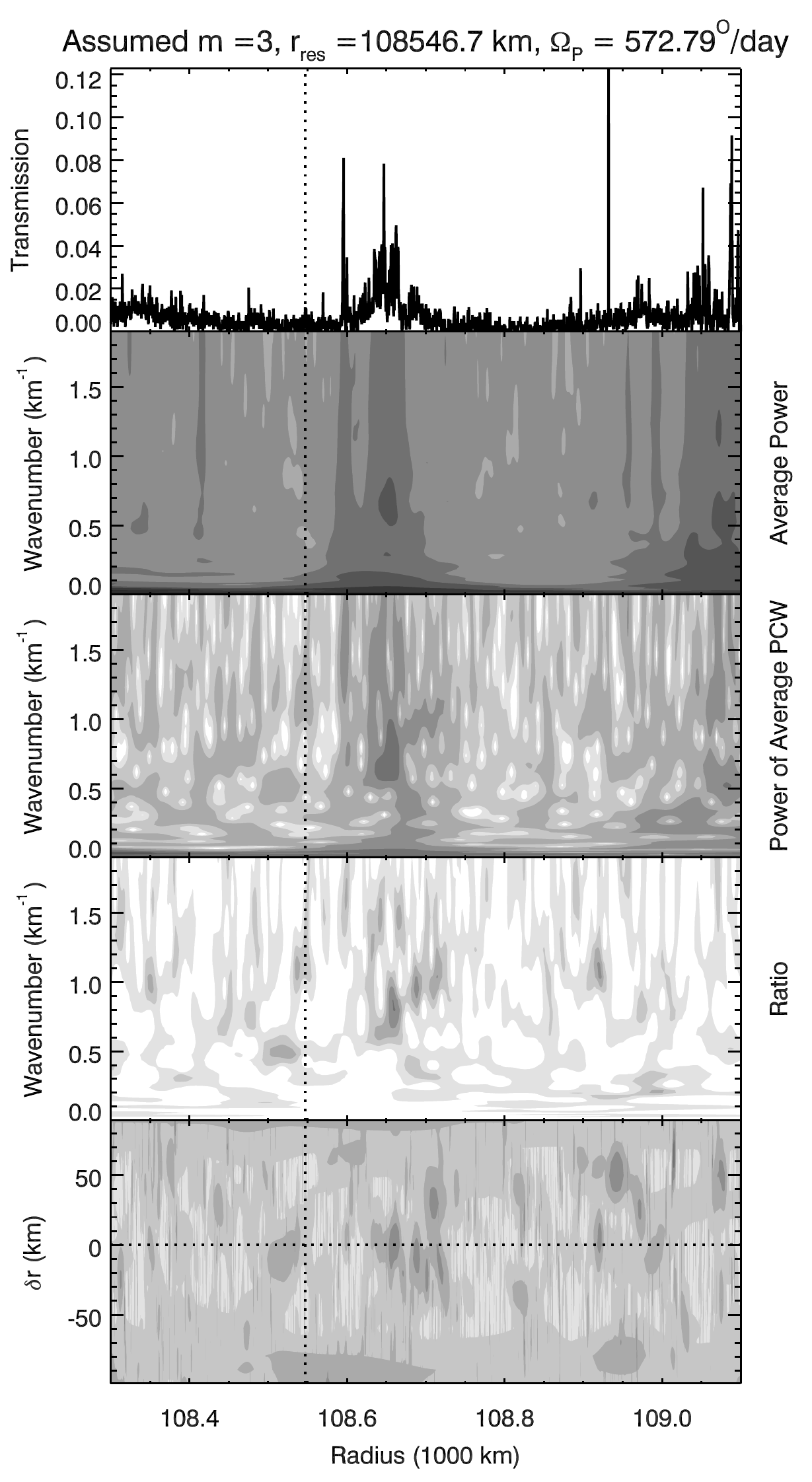}}}
\caption {Wavelet analysis on the Pandora 3:2 wave in the same format as Figure~\ref{pr76}.
A wave-like signature can be observed between 108,600 and 108,700 km in the ratio plot. This signal is strongest where $\delta r=0$, i.e. when the assumed pattern speed is close to the expected pattern speed for this density wave.}
\label{pd32wave}
\end{figure}

We consider the remaining potential wave signatures (associated with the Enceladus 3:1, Janus 3:2 and Pandora 3:2 resonances) as a group because all three of these features occupy regions of high opacity  and large stochastic variations in optical depth that obscure the relevant wave signals in individual profiles. Figures~\ref{en31wave}-\ref{pd32wave} show the results of the phase-corrected wavelet analysis for these waves. In all three cases the average wavelet power shows signal over a wide range of wavenumbers wherever the transmission is finite. These are due to the stochastic variations in the ring's transmission, which are suppressed in the average phase-corrected wavelet.
Indeed, both the power of the average  phase-corrected wavelet and the power ratios reveal potential density wave signatures. 

The Enceladus 3:1 resonance provides the most compelling evidence for a previously undetected density wave, with the phase-corrected wavelet power showing a discontinuous diagonal linear band  extending from 115,300 km to 115,650 km and ranging over wavenumbers between 0.4 km$^{-1}$ and 1.2 km$^{-1}$ (see Figure~\ref{en31wave}).  This trend is consistent with the expected signal from a density wave, and  extrapolating this trend inwards would produce an intercept close to the expected resonance location around 115,200 km. Furthermore, this signal is only apparent when the assumed pattern speed approximately matches the expected pattern speed for the Enceladus 3:1 resonance (see bottom panel of Figure~\ref{en31wave}). Hence it is reasonable to conclude that this is indeed the signature of a density wave.

For the Janus 3:2 resonance, the average phase-corrected wavelet contains a region of strong periodic signals at wavenumbers around 0.5 km$^{-1}$ near 116,170 km, and a slightly weaker signal at wavenumbers around 0.2 km$^{-1}$ and radii of 116,050 km (see Figure~\ref{ja32wave}).  These two regions of enhanced signal follow a similar diagonal trend as the bands seen in the above density waves, and extrapolating this trend inwards would produce an intercept close to the expected resonance location just interior to 116,000 km. Furthermore, these signals are only present when the assumed pattern speed is close to the expected pattern speeds of the Janus 3:2 resonance (see bottom panel of Figure~\ref{ja32wave}), consistent with the signature of the appropriate density wave. Note that the Janus 3:2 resonance is comparable in strength to the Janus 2:1 wave discussed above, so the comparative weakness of this signal must be due to either the ring's substantially higher optical depth or to interference by other fine-scale structure.

Finally, for the Pandora 3:2 resonance there is a very weak signal between 0.5 km$^{-1}$ and 1.0 km$^{-1}$ and between 108,600 km  and 108,700 km that could represent a density wave showing through a narrow region of finite optical depth in the B-ring's core (see Figure~\ref{pd32wave}). Again, these enhancements are only detectable when the pattern speed is close to the appropriate resonance pattern speed. Hence even here we have evidence for a wave signature in the occultation data.

\begin{figure}
\centerline{\resizebox{3in}{!}{\includegraphics{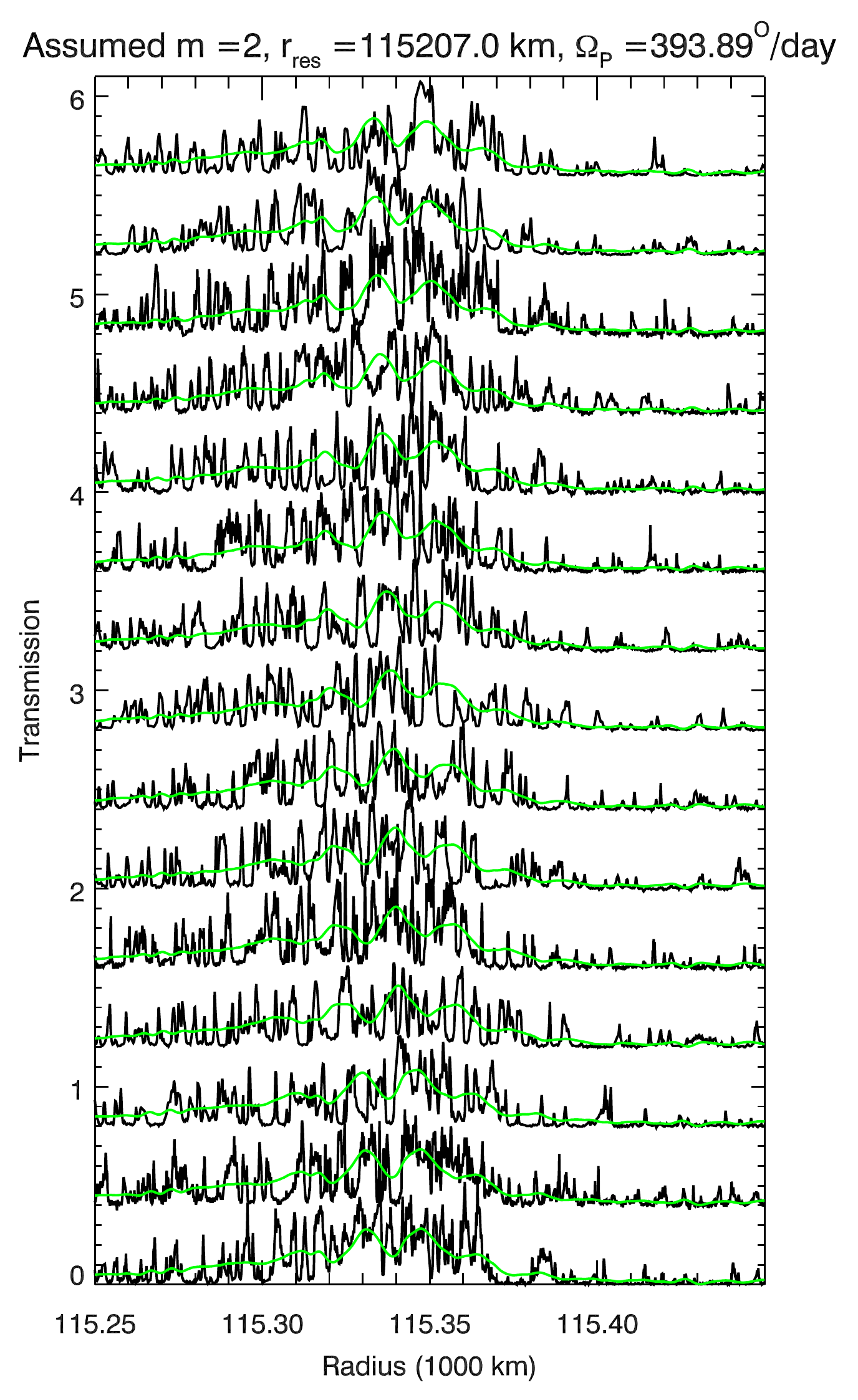}}}
\caption{Profiles of the region around the Enceladus 3:1 wave. The black profiles are the observed data, offset for clarity and ordered by the predicted wave phase $\phi_{\lambda t}$. The overlaid green curves are the reconstructed wave signal derived from the phase-corrected average wavelet between wavelengths between 5 and 100 km. The periodic signatures in the reconstructed profiles have been scaled up by a factor of four for the sake of clarity, and each curve has had the wave's phase adjusted to match its predicted value for the relevant profile.}
\label{en31profs}
\end{figure}

\begin{figure}
\centerline{\resizebox{3in}{!}{\includegraphics{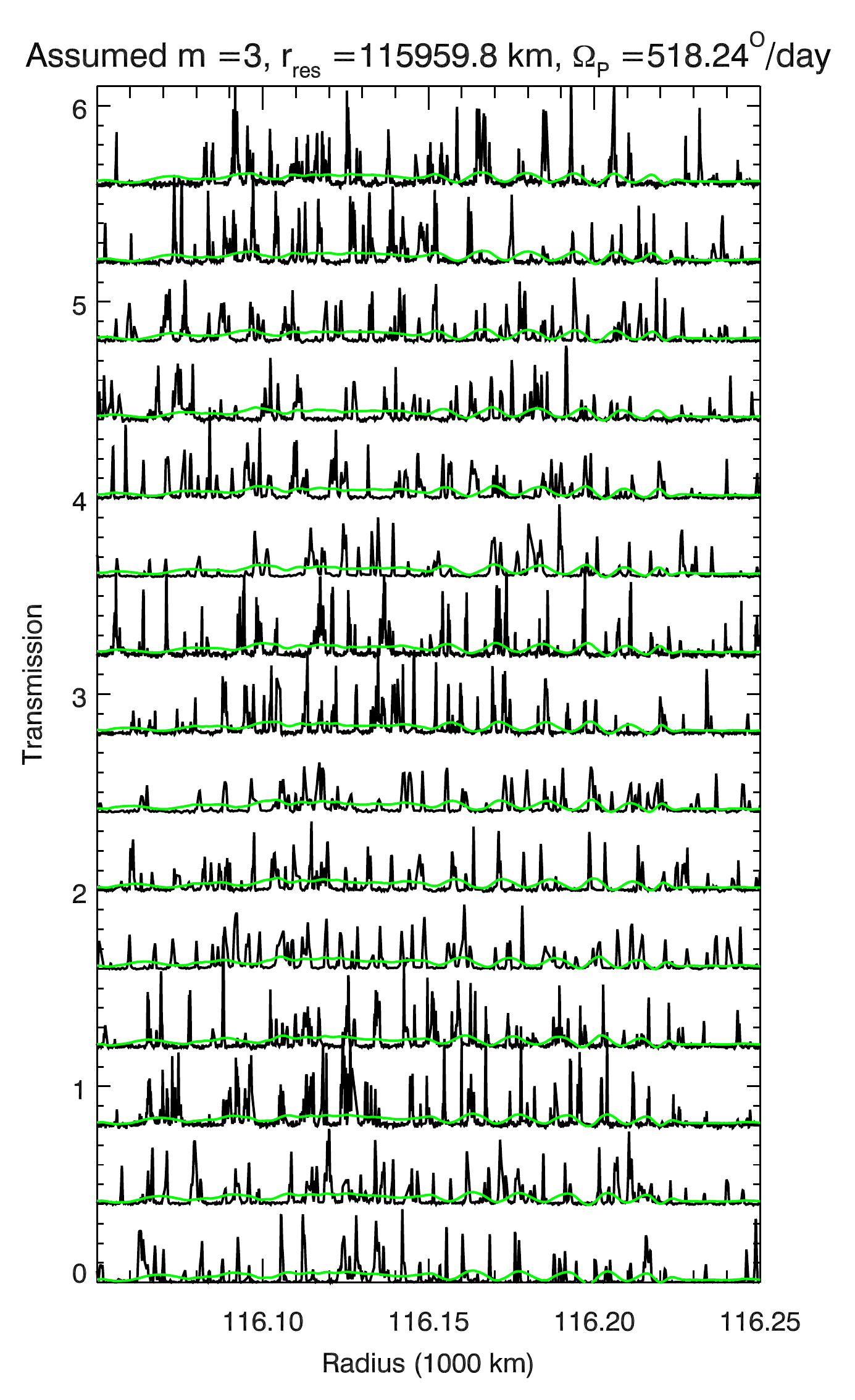}}}
\caption{Profiles of the region around the Janus 3:2 wave. The black profiles are the observed data, offset for clarity and ordered by the predicted wave phase $\phi_{\lambda t}$. The overlaid green curves are the reconstructed wave signal derived from the phase-corrected average wavelet between wavelengths between 5 and 100 km. The periodic signatures in the reconstructed profiles have been scaled up by a factor of four for the sake of clarity, and each curve has had the wave's phase adjusted to match its predicted value for the relevant profile.}
\label{ja32profs}
\end{figure}

\begin{figure}
\centerline{\resizebox{3in}{!}{\includegraphics{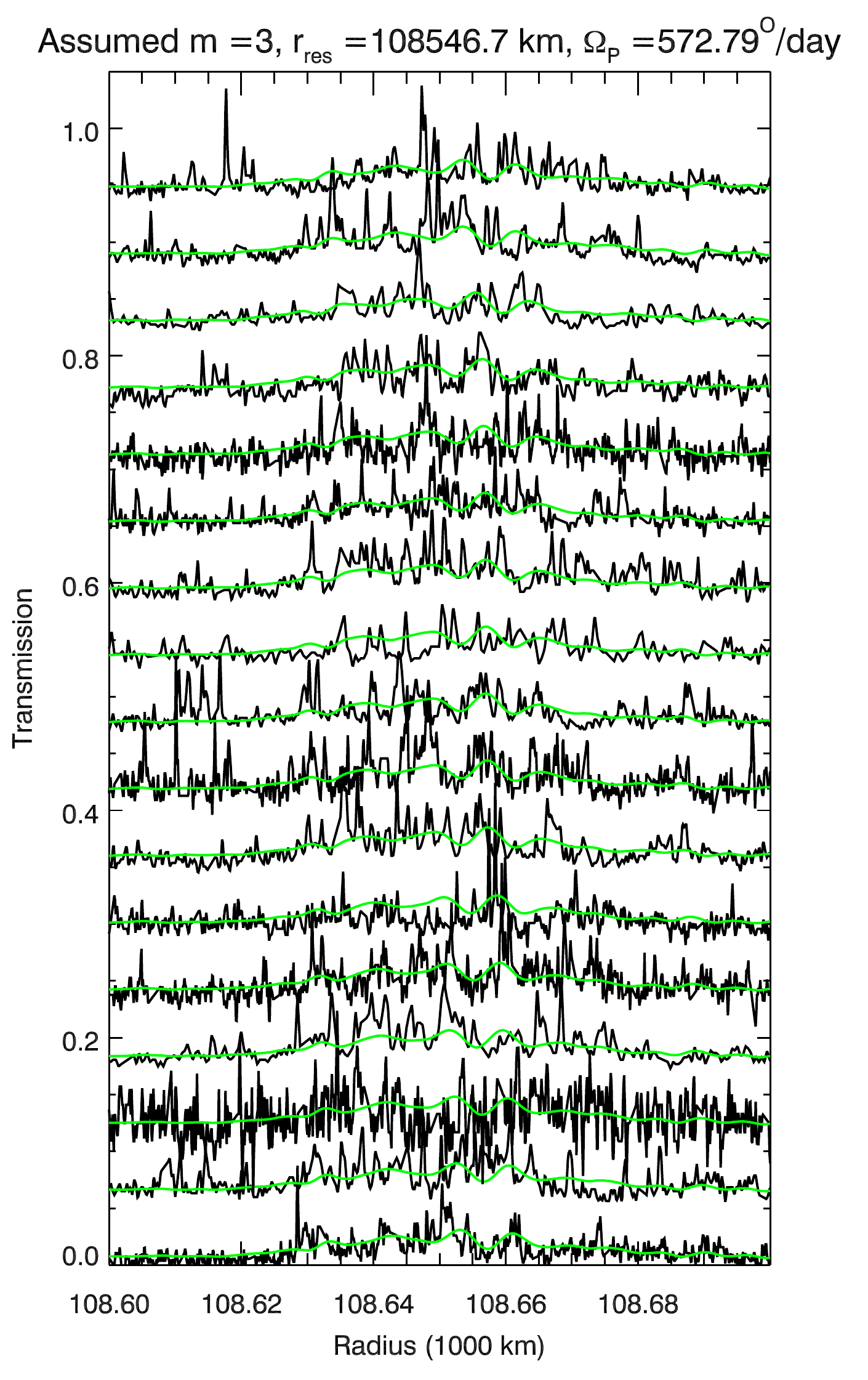}}}
\caption{Profiles of the region around the Pandora 3:2 wave. The black profiles are the observed data, offset for clarity and ordered by the predicted wave phase $\phi_{\lambda t}$. The overlaid green curves are the reconstructed wave signal derived from the phase-corrected average wavelet between wavelengths between 5 and 100 km. The periodic signatures in the reconstructed profiles have been scaled up  by a factor of four for the sake of clarity,  and each curve has had the wave's phase adjusted to match its predicted value for the relevant profile.}
\label{pd32profs}
\end{figure}

To illustrate why multiple occultation profiles are essential for identifying these very weak wave signals, Figures~\ref{en31profs}-\ref{pd32profs} show the observed profiles, sorted by the relevant phase parameter. Overplotted in green on each profile is the signal recovered from the combined phase-corrected wavelet (using wavelengths between 5 km and 100 km), which represents the opacity variations that would be due to the wave signal alone in each profile (including the appropriate phase shifts). While there are some correlations between the observed profiles and the recovered wave signal (e.g. the region around 115,350 km for the Enceladus 3:1 wave and the region around 108,650 km for the Pandora 3:2 wave), it is very difficult to identify convincing wave signatures in individual profiles due to the other opacity variations superimposed on top of the wave. Hence we cannot confirm that these wave signatures are real by simple inspection of the profiles. Nevertheless, since these wave-like  signatures only appear when the appropriate pattern speeds for the relevant resonances are used, it is reasonable to conclude that these are real density wave signals.

\begin{figure}
\centerline{\resizebox{4in}{!}{\includegraphics{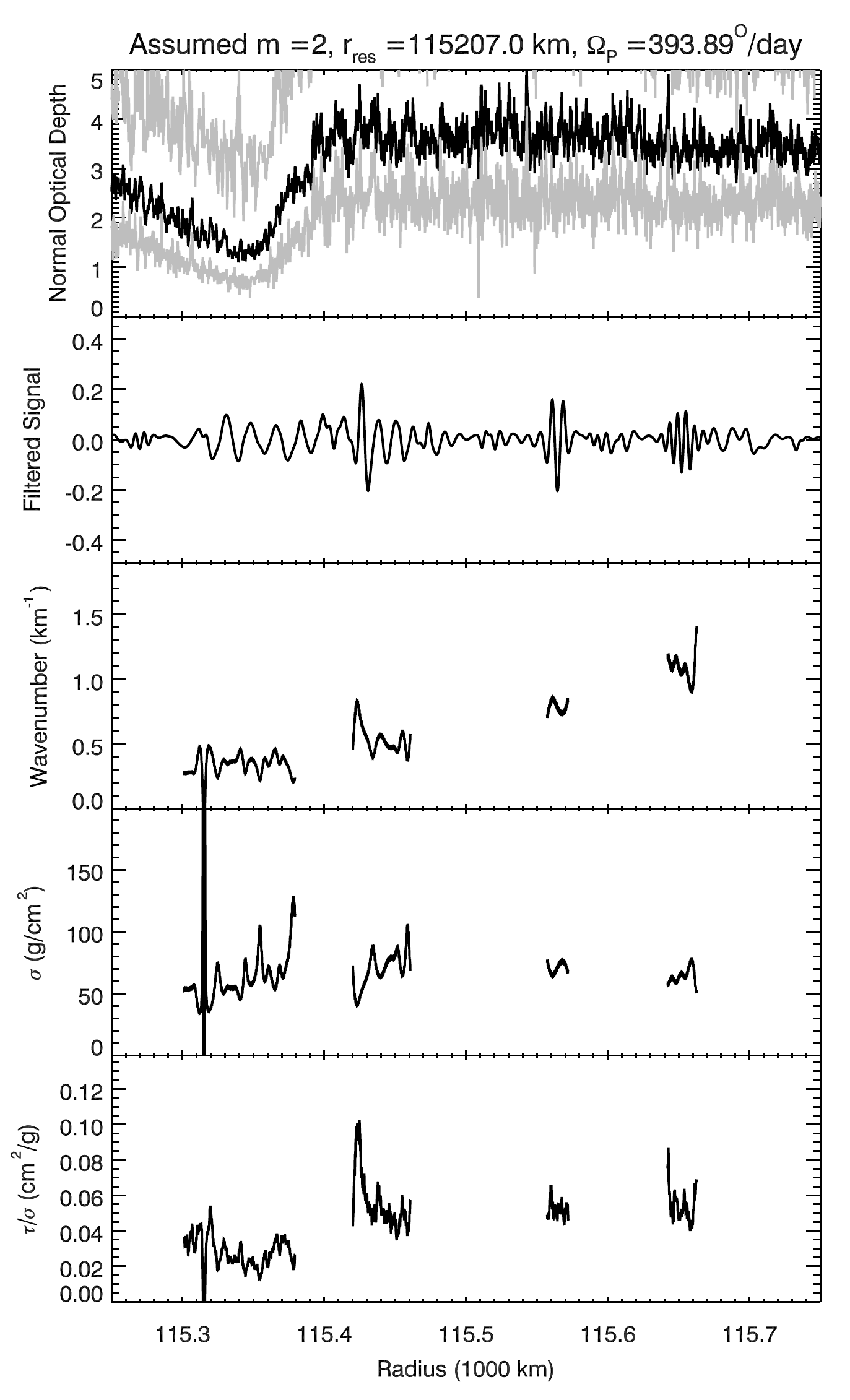}}}
\caption{Parameters for the Enceladus 3:1 wave derived from the average phase-corrected wavelet, in the same format as Figure~\ref{pr76m}. In this case the reconstructed profile is derived from the average phase-corrected wavelet data for wavelengths between 5 and 100 km, and only data where the peak power ratio was above 0.3 are shown in the bottom three panels.} 
\label{en31m}
\end{figure}

\begin{figure}
\centerline{\resizebox{4in}{!}{\includegraphics{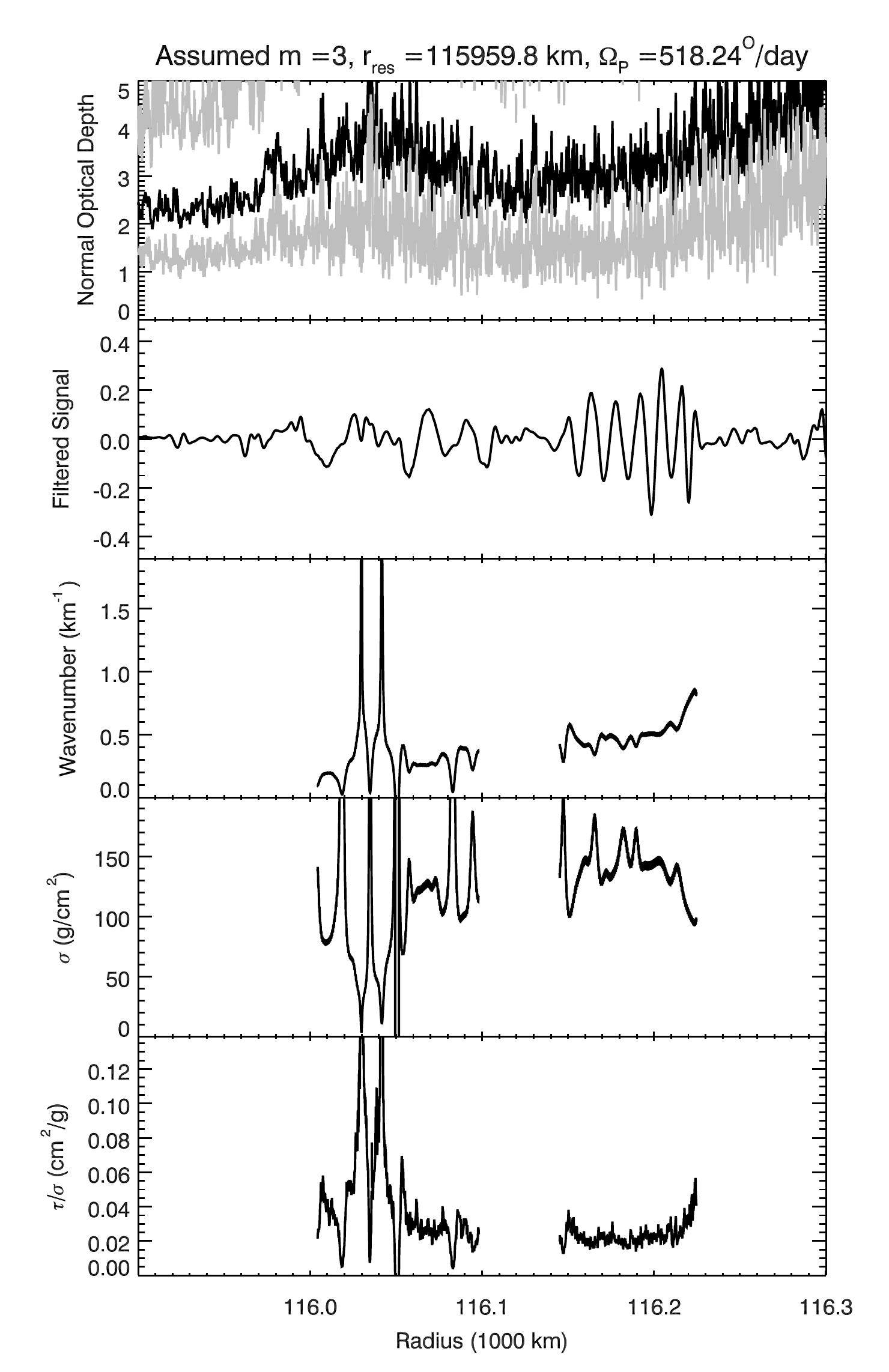}}}
\caption{Parameters for the Janus 3:2 wave derived from the average phase-corrected wavelet, in the same format as Figure~\ref{pr76m}. In this case the reconstructed profile is derived from the average phase-corrected wavelet data for wavelengths between 5 and 100 km, and only data where the peak power ratio was above 0.3 are shown in the bottom three panels.}
\label{ja32m}
\end{figure}

\begin{figure}
\centerline{\resizebox{4in}{!}{\includegraphics{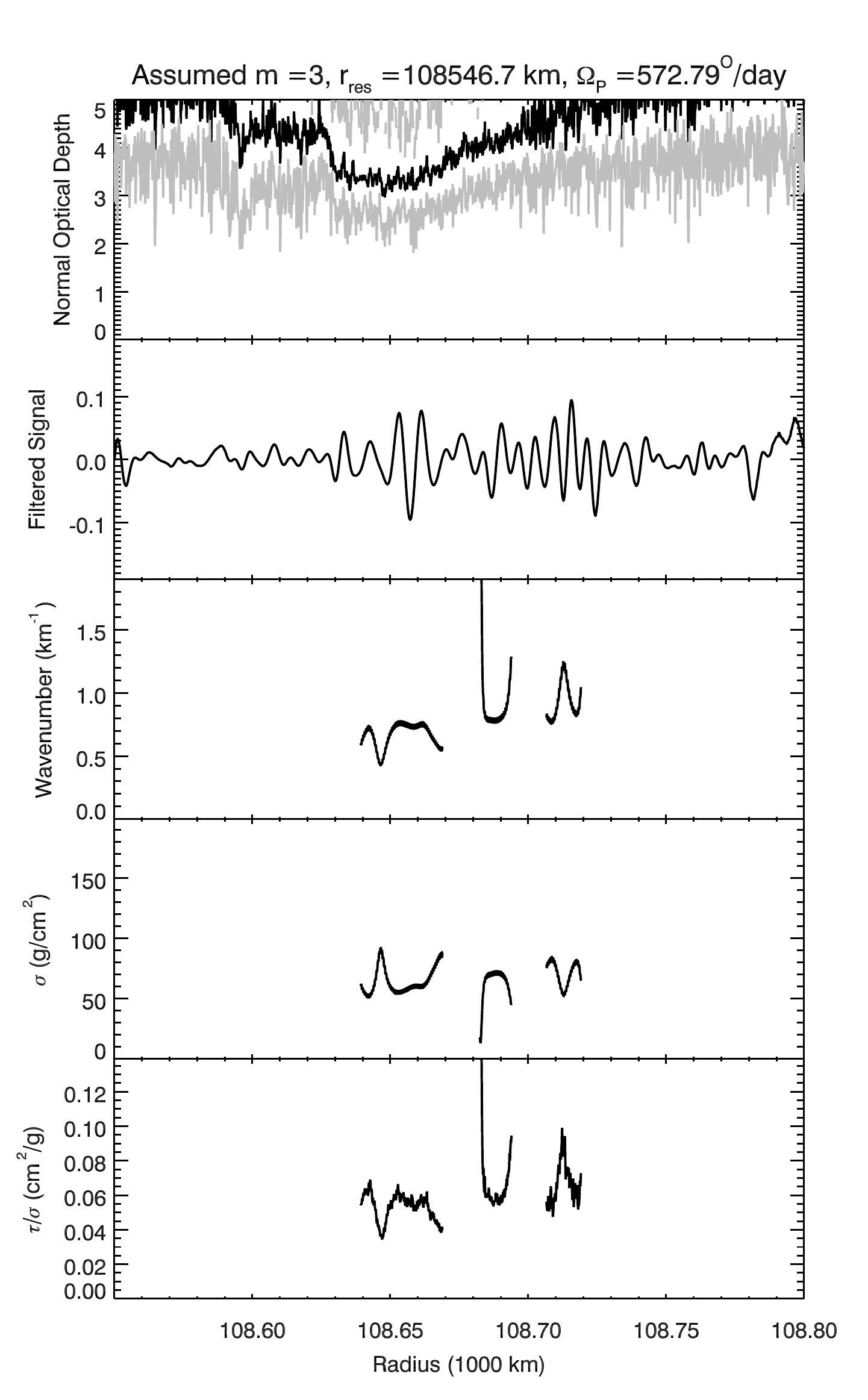}}}
\caption{Parameters for the Pandora 3:2 wave derived from the average phase-corrected wavelet, in the same format as Figure~\ref{pr76m}. In this case the reconstructed profile is derived from the average phase-corrected wavelet data for wavelengths between 5 and 100 km, and only data where the peak power ratio was above 0.3 are shown in the bottom three panels.}
\label{pd32m}
\end{figure}

Assuming that all these signals are indeed real wave signatures, we can use the same procedures described in Section~\ref{wavelets} above to obtain reconstructed wave profiles from the average phase-corrected wavelet (in these cases, we integrate over wavenumbers between $2\pi/5$ km and $2\pi/ 100$ km), and then extract estimates of the patterns' wavenumber, as well as the inferred surface mass density of the ring. The results of these calculations are shown in Figures~\ref{en31m}-\ref{pd32m}. For all three regions, we see the wave-like signature has a variable amplitude, consistent with the patchy signal in the average phase-corrected wavelet. Nevertheless, where the signal is clear (i.e. where the peak power ratio is greater than 0.3), the wavenumber does show a reasonably monotonic trend with radius. 

The mass density and $\tau_n/\sigma$ estimates derived from these features are also reasonably consistent for each wave.  For the Janus 3:2 wave,  we can see the wave-like signal in two regions. In the outer region centered around 116,185 km, the wave is consistent with a local surface mass density around 140 g/cm$^2$. For the inner region, the mass density estimates are erratic interior to 116,060 km, but gives stable estimates around 140 g/cm$^2$ between 116,060 km and 116,080 km.
 For the Enceladus 3:1 wave, we have four discrete regions that yield sensible periodic wave signals, and all are consistent with mass densities around 70 g/cm$^2$.  A similar range of mass densities are derived for the three regions where the Pandora 3:2 wave signal is sufficiently strong to detect. Thus these new mass density estimates seem to be internally consistent within each wave.

\FloatBarrier

\section{Discussion}
\label{discussion}

\begin{table}
\caption{Estimates of the average normal optical depth and surface mass density derived from waves in the B ring.}
\label{masstab}
\begin{tabular}{|c|c|c|c|c|c|}\hline
Wave & Radial	Range &	Mean Rad. & Mean $\tau_n$ & Mean $\sigma$ & Mean $\tau_n/\sigma$ \\
&  (km) & (km) &	& (g/cm$^2$) & (cm$^2$/g) \\ \hline 
&  96300.-  96400. &   96350. & 1.08$\pm$0.08 &  69.03$\pm$  5.98 & 0.0157$\pm$0.0015 \\
&  96475.-  96494. &   96485. & 1.69$\pm$0.12 &  58.65$\pm$  7.93 & 0.0292$\pm$0.0040 \\
&  96521.-  96538. &   96530. & 1.24$\pm$0.09 &  46.72$\pm$  5.58 & 0.0271$\pm$0.0046 \\
 Janus &  96558.-  96572. &   96565. & 1.65$\pm$0.10 &  46.11$\pm$  3.67 & 0.0361$\pm$0.0041 \\
 2:1 & 96602.-  96640. &   96610. & 1.95$\pm$0.11 &  47.03$\pm$  3.81 & 0.0417$\pm$0.0036 \\
 & 96640.-  96656. &   96647. & 1.89$\pm$0.08 &  44.84$\pm$  6.07 & 0.0429$\pm$0.0064 \\
 & 96680.-  96692. &   96686. & 1.97$\pm$0.09 &  46.69$\pm$  3.90 & 0.0424$\pm$0.0040 \\
 \hline
Mimas & 101390.- 101480. &  101435. & 1.90$\pm$0.27 &  42.07$\pm$  4.77 & 0.0456$\pm$0.0093 \\
5:2 & 101550.- 101600. &  101575. & 1.40$\pm$0.03 &  39.10$\pm$  1.49 & 0.0358$\pm$0.0015 \\
\hline
& 108640.- 108669. &  108654. & 3.36$\pm$0.15 &  64.95$\pm$ 10.76 & 0.0529$\pm$0.0078 \\
Pandora 3:1 & 108640.- 108699. &  108654. & 4.15$\pm$0.15 &  67.24$\pm$ 6.82 & 0.0626$\pm$0.0094 \\
& 108707.- 108719. &  108713. & 4.68$\pm$0.34 &  71.60$\pm$  9.73 & 0.0668$\pm$0.0118 \\
\hline
& 115330.- 115379. &  115355. & 1.68$\pm$0.48 &  67.71$\pm$ 17.68 & 0.0254$\pm$0.0061 \\
Enceladus &  115420.- 115461. &  115440. & 3.72$\pm$0.34 &  70.27$\pm$ 14.06 & 0.0556$\pm$0.0151 \\
3:1 & 115557.- 115572. &  115565. & 3.70$\pm$0.30 &  71.05$\pm$  4.01 & 0.0522$\pm$0.0045 \\
& 115642.- 115663. &  115652. & 3.43$\pm$0.34 &  64.51$\pm$  5.98 & 0.0538$\pm$0.0083 \\
\hline
Janus & 116060.- 116080. &  116070. & 3.31$\pm$0.51 & 120.09$\pm$ 8.42 & 0.0277$\pm$0.0047 \\
3:2 & 116150.- 116220. &  116185. & 3.14$\pm$0.40 & 141.02$\pm$ 16.38 & 0.0226$\pm$0.0042 \\
\hline
Mimas 4:2 & 116500.- 116750. &  116625. & 1.80$\pm$0.49 &  54.00$\pm$ 10.00$^a$ & 0.0334$\pm$0.0096 \\ \hline\end{tabular}

 $^a$ from \citet{Lissauer85}
\end{table}

\begin{figure}
\centerline{\resizebox{4in}{!}{\includegraphics{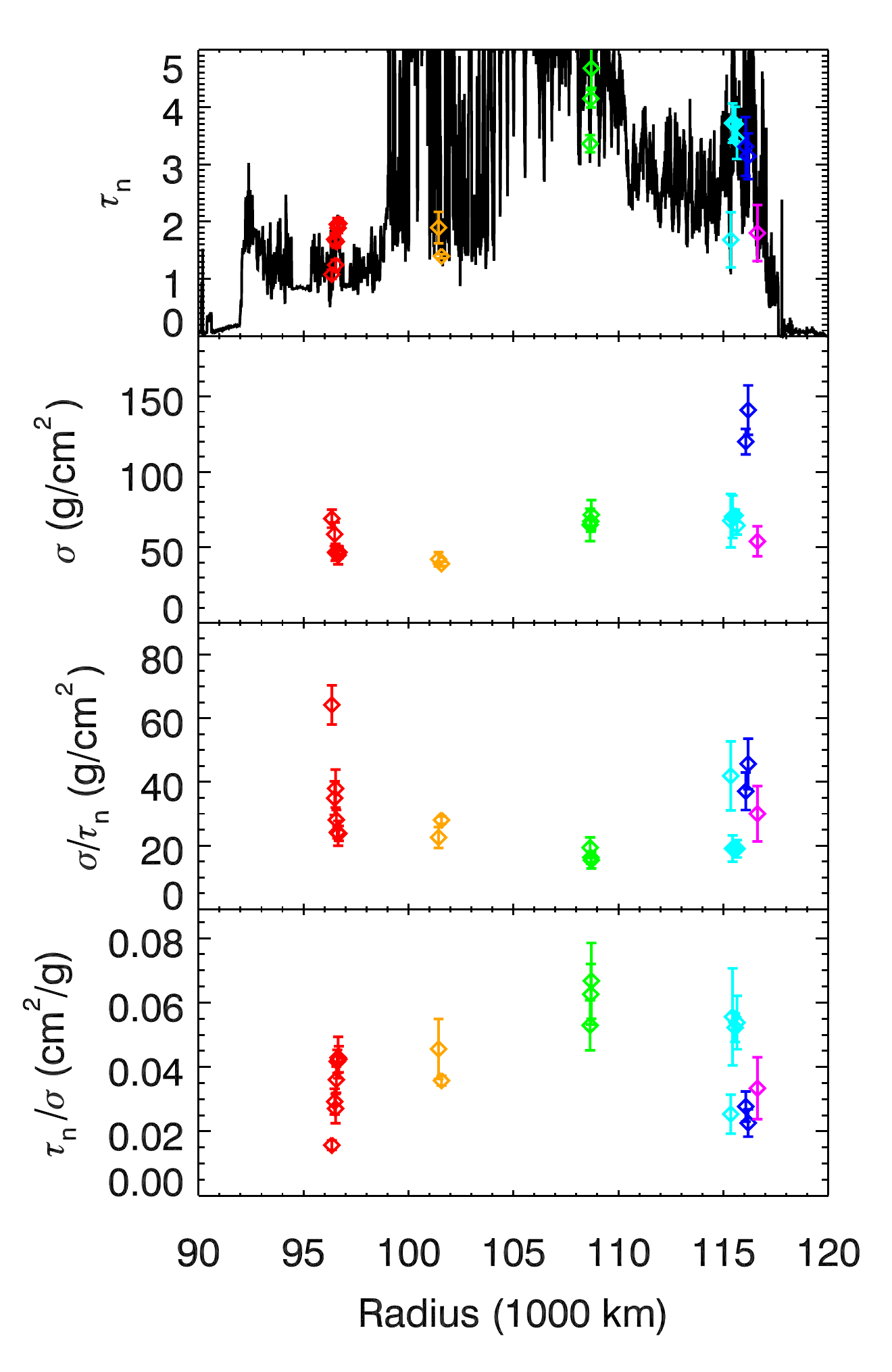}}}
\caption{Summary of mass density and opacity estimates of Saturn's B ring. The panels show the optical depth $\tau_n$, surface mass density $\sigma$, the ratio $\sigma/\tau_n$ and its reciprocal as functions of radius across the B ring. Note that for a wide range of optical depths the surface mass density appears to remain around 70 g/cm$^2$.}
\label{massrad}
\end{figure}

\begin{figure}
\centerline{\resizebox{4in}{!}{\includegraphics{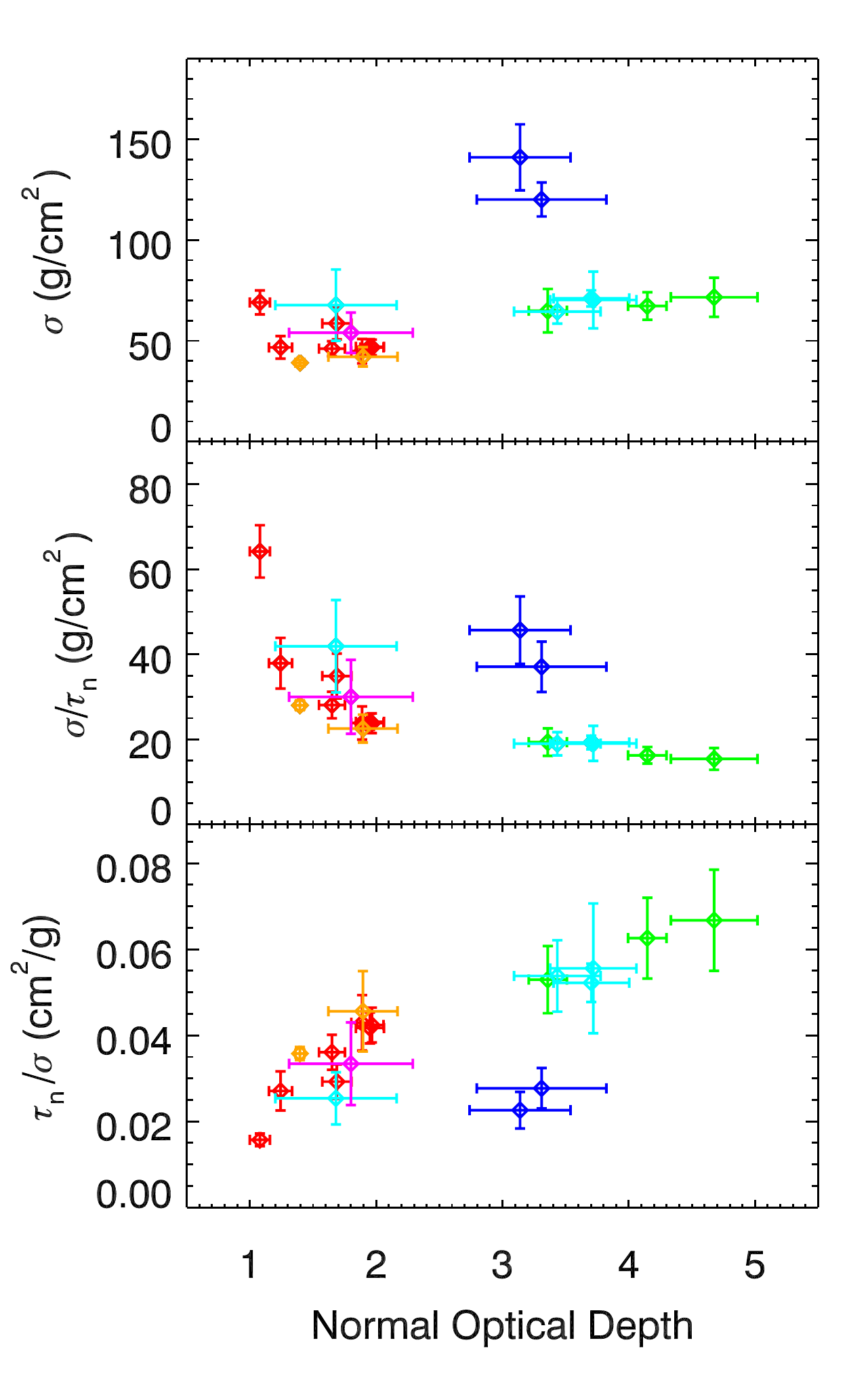}}}
\caption{The B-ring surface mass density $\sigma$, $\sigma/\tau_n$ and its reciprocal as functions of the normal optical depth. Note the different colored points correspond to different waves, using the same color codes as Figure~\ref{massrad}.}
\label{masstau}
\end{figure}

\begin{figure}
\centerline{\resizebox{4in}{!}{\includegraphics{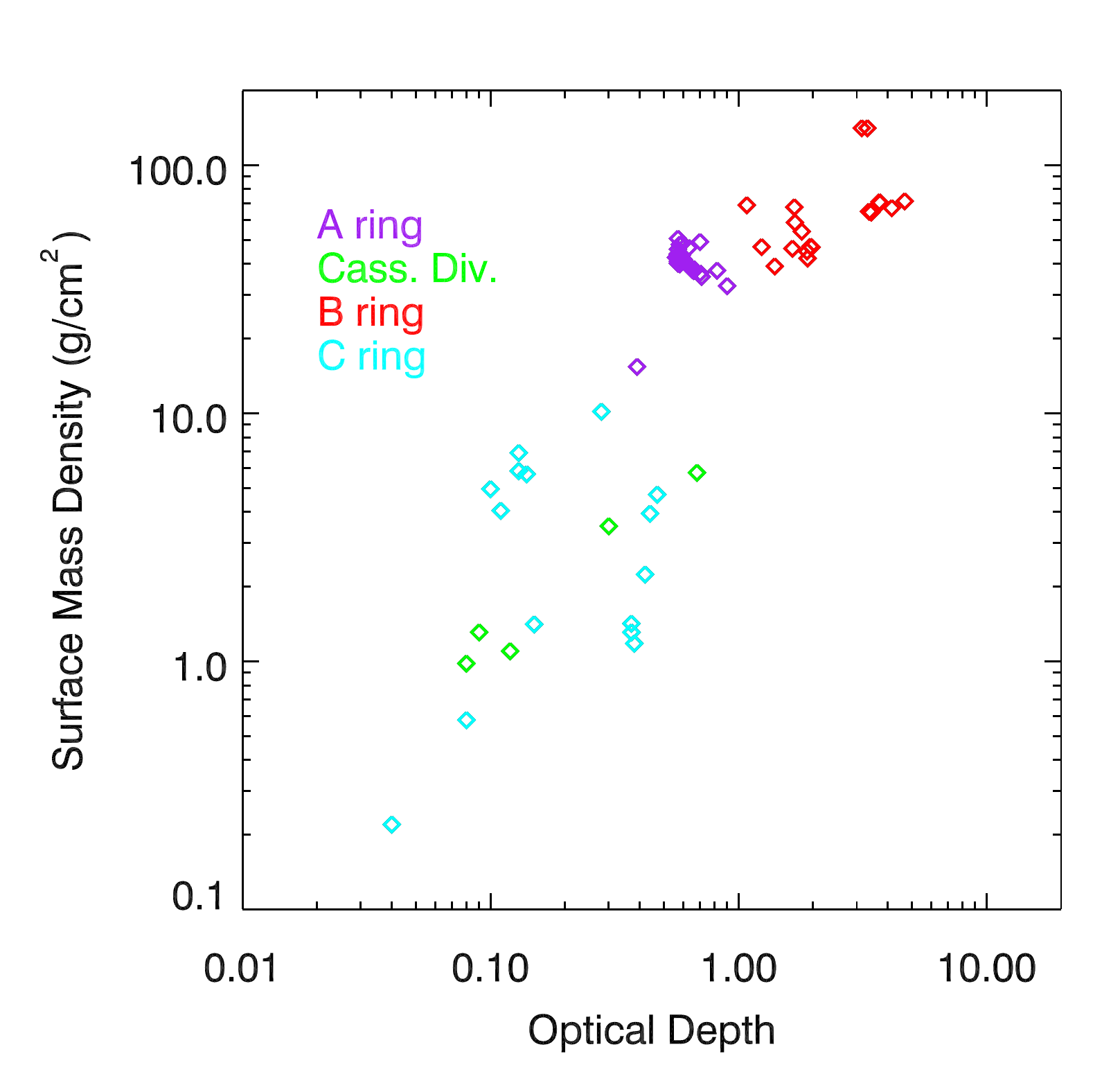}}}
\caption{Plot of the ring's surface mass density versus normal optical depth for all of Saturn's rings, based on density wave analyses from \citet{Lissauer85, Tiscareno07, Colwell09, Baillie11, HN14} and this work. Note the broad range of optical depths possible for any given mass density.}
\label{tausigma}
\end{figure}

Table~~\ref{masstab} and Figure~\ref{massrad} summarize the mass density estimates derived from this analysis. For each density wave, we derive separate mass densities from each part of the wave where the signal is clear, and for the Janus waves we only consider regions where the pattern speed is close to Janus' mean motion. We also include the surface mass density derived from the Mimas 4:2 bending wave reported by \citet{Lissauer85} for reference. What is most remarkable about all these new estimates is that the derived mass densities are generally quite low, with the Janus 3:2 wave yielding values around 140 g/cm$^2$ and all the other wave signatures giving between 40 and 70 g/cm$^2$, within a factor of two of the A-ring's typical surface mass density.

Another striking aspect of these measurements is that the B-ring's optical depth seems to be largely uncorrelated with its mass density. For example, the high mass density derived from the Janus 3:2 wave does not correspond to the most opaque region. Furthermore, the other five waves yield similar mass densities for regions with optical depths ranging from $\sim1.5$ to almost  5. Indeed Figure~\ref{masstau} shows no obvious trend in the mass density versus optical depth within these waves. This result, while surprising, is consistent with recent analyses of density waves elsewhere in Saturn's rings.  \citet{Tiscareno13} found that the sharp rise in optical depth near the A-ring's inner edge did not correspond to a marked jump in mass density, while \citet{Baillie11} and \citet{HN14} showed that even though the C-ring plateaux are several times more opaque than the background C ring, these two regions have nearly the same mass density. Indeed, as shown in Figure~\ref{tausigma}, it appears that rings with a given mass density can have optical depths that vary by almost an order of magnitude. {\sl Optical depth therefore cannot be regarded as a reliable proxy for the ring's mass density in any part of Saturn's rings}.

These findings also have implications for the total mass of Saturn's rings. Recent work has considered the possibility that the B-ring's high opacity might require extremely high surface mass densities \citep{Robbins10, Charnoz11, Hedman13}, but these new measurements suggest that this may not be justified. Instead, the total mass of the B ring could be quite low, which may be consistent with estimates based on the charged-particle populations near the rings \citep{Cooper85}\footnote{It should be noted that this work yielded two different estimates of the mass density. The authors favored the solution with a surface mass density around 70 g/cm$^2$, but could not exclude another solution with a value of around 350 g/cm$^2$} and the high porosity of the ring particles inferred from thermal infrared data \citep{Reffet15}.

The B-ring's mass is most easily compared to the masses of Mimas (the smallest of Saturn's mid-sized quasi-spherical satellites) and the A ring. The mass of Mimas has been well measured by its gravitational interactions with other moons, and is $M_M=3.7493\pm0.0031\times10^{19}$ kg \citep{Jacobson06}.
The total mass of the A ring is also fairly well constrained because it contains many density waves, and aside from a few narrow gaps, the ring does not show strong variations in its surface mass density and optical depth between 122,357 km and its outer edge at 136,780 km.\footnote{Note the classical inner edge of the A ring is at 122,050 km, but \citet{Tiscareno13} showed the ring's mass density only reaches typical A-ring values at 122,357 km.} The A ring therefore covers about $1.2\times10^{10}$ km$^2$, and its average surface mass density is between 35 and 40 g/cm$^2$ \citep{Tiscareno07}, and so the A-ring's total mass is between 4 and $5\times10^{18}$ kg, or 0.11-0.13 $M_M$.  

Since the B-ring's optical depth cannot provide a reliable estimate of its mass density, the limited number of density waves examined here do not allow us to derive very precise estimates of the B-ring's total mass. However, these few measurements do indicate that the B-ring's total mass is substantially less than that of Mimas. The B ring covers a surface area of about $1.7\times10^{10}$ km$^2$, so in order for the B-ring's mass to equal the mass of Mimas, it would have to have an average surface mass density of about 220 g/cm$^2$, or an average $\tau_n/\sigma$ less than 0.013 cm$^2$/g. The required mass density is well above any of our estimates, and the required $\tau_n/\sigma$ is lower than any of our numbers. Indeed, if we neglect the relatively high mass density estimates from the Janus 3:2 wave, then the remaining data points indicate the B-ring's mass density is between 45 and 70 g/cm$^2$. This would imply the B-ring's total mass  is $7-12\times10^{18}$ kg or 0.20-0.32 $M_M$, which is only two or three times the A-ring's mass.  Even if we take the Janus 3:2 mass density as an upper limit, this only yields a mass of $24\times10^{18}$ kg, or about 0.68 $M_M$. It therefore appears likely that the B ring's total mass is between one-third and two-thirds of Mimas' mass, comparable to early Voyager-based estimates \citep{Esposito83, Cooper85}.

Of course, this new estimate of the ring's total mass is based on just a few locations in the B ring, and so one could argue that most of the B-ring's mass is hidden in the truly opaque parts of the rings. However, the measurements considered here include regions with optical depths between 3.5 and 4.5, so these massive regions would have to be those with opacities above 4 or 5.   Note that only about 27\% of the ring's surface area has an optical depth greater than 4, and less than 16\% has an optical depth above 5, so there is not much space to hide a large amount of mass. Alternatively, one could suggest that in high-optical-depth regions waves do not propagate in accordance with the standard theories, causing the value of $\sigma$ derived from Equation~\ref{keq} to be biased low. This also seems to be rather unlikely, however, given the relative consistency in the mass densities derived from the different parts of each wave. Fortunately, Cassini will perform experiments at the end of its mission in 2017 that should provide independent constraints on the ring's total mass \citep{SB09, Spilker14} that can confirm or refute the results of this study.

{\bf Acknowledgements:} This work was supported in part by the NASA Cassini  Data Analysis Program grant NNX14AD50G. We also wish to thank the VIMS planning team and the Cassini Project for providing the data used in this analysis. We want to thank M.S. Tiscareno for useful discussions regarding wavelet transforms, as well as J. Schmidt and an anonymous reviewer for their helpful comments and positive reviews on an earlier version of this manuscript.

\pagebreak


\end{document}